\documentclass[12pt]{article}
\usepackage{amsmath}

\usepackage{setspace}
\usepackage[super,compress]{cite}
\usepackage{graphicx}
\usepackage[polutonikogreek,english]{babel}
\usepackage{comment}
\renewcommand{\d}{{\rm d}}
\newcommand{\rl}{\ell}
\newcommand{\tilS}{\tilde{S}}
\newcommand{\bN}{\mbox{\boldmath$N$}}
\newcommand{\tg}{{\widetilde{g}}}
\newcommand{\trv}{{\widetilde{\mbox{\rm v}}}}
\newcommand{\tl}{\widetilde{l}}
\newcommand{\rv}{\mbox{\rm v}}
\newcommand{\tf}{{\widetilde{f}}}
\newcommand{\tih}{{\widetilde{h}}}
\newcommand{\bv}{\mbox{\boldmath$v$}}
\newcommand{\pR}{\,^{+}\!R}
\newcommand{\pOmega}{\,^+\!\Omega}
\newcommand{\mOmega}{\,^-\!\Omega}
\newcommand{\bl}{\mbox{\boldmath$l$}}
\newcommand{\pmOmega}{\,^\pm\!\Omega}
\newcommand{\pmR}{\,^{\pm}\!R}
\newcommand{\overOm}{\overline{\Omega}}
\newcommand{\overT}{\overline{T}}
\newcommand{\bomega}{\mbox{\boldmath$\omega$}}
\newcommand{\be}{\mbox{\boldmath$e$}}
\newcommand{\N}{{\cal N}}
\newcommand{\sh}{\mathop{\rm sh}\nolimits}
\newcommand{\cF}{{\cal F}}
\newcommand{\teta}{{\widetilde{\eta}}}
\newcommand{\otheta}{\overline{\vartheta}}
\newcommand{\rmm}{\mbox{\rm m}}
\newcommand{\rt}{\mbox{\rm t}}
\newcommand{\hlambda}{{\widehat{\lambda}}}
\newcommand{\no}{<}
\newcommand{\hmu}{{\widehat{\mu}}}
\newcommand{\trmm}{{\widetilde{\rmm}}}
\newcommand{\trt}{{\widetilde{\rt}}}
\newcommand{\hone}{{\widehat{1}}}
\newcommand{\htwo}{{\widehat{2}}}
\newcommand{\halpha}{{\widehat{\alpha}}}
\newcommand{\hthr}{{\widehat{3}}}
\newcommand{\hast}{{\widehat{\ast}}}
\newcommand{\hnul}{{\widehat{0}}}
\newcommand{\hbeta}{{\widehat{\beta}}}
\newcommand{\tk}{{\widetilde{k}}}
\newcommand{\ttrt}{{\widetilde{\widetilde{\rt}}}}
\newcommand{\ttrmm}{{\widetilde{\widetilde{\rmm}}}}
\newcommand{\rg}{\mbox{\rm g}}
\newcommand{\srg}{{\rm g}}
\newcommand{\tDelta}{{\widetilde{\Delta}}}
\newcommand{\otDelta}{{\overline{\widetilde{\Delta}}}}
\newcommand{\oDelta}{{\overline{\Delta}}}
\newcommand{\oT}{{\overline{T}}}
\newcommand{\rtheta}{{\rm \textgreek{j}}}
\newcommand{\ortheta}{\overline{\rm \textgreek{j}}}

\newcommand{\bp}{{\mbox{\boldmath$p$}}}
\newcommand{\bq}{{\mbox{\boldmath$q$}}}
\newcommand{\sbp}{{\mbox{\scriptsize \boldmath$p$}}}
\newcommand{\sbq}{{\mbox{\scriptsize \boldmath$q$}}}
\renewcommand{\Re}{\mathop{\rm Re}\nolimits}
\renewcommand{\Im}{\mathop{\rm Im}\nolimits}
\newcommand{\tp}{{\widetilde{p}}}
\newcommand{\tr}{{\widetilde{r}}}
\newcommand{\tbp}{{\widetilde{\bp}}}
\newcommand{\tq}{{\widetilde{q}}}
\newcommand{\tbq}{{\widetilde{\bq}}}
\newcommand{\ts}{{\widetilde{s}}}

\newcommand{\rh}{\mbox{\rm h}}
\newcommand{\srh}{{\rm h}}

\allowdisplaybreaks[4]

\begin{document}

\title{On the gravitational diagram technique in the discrete setup}

\author{V.M. Khatsymovsky \\
 {\em Budker Institute of Nuclear Physics} \\ {\em of Siberian Branch Russian Academy of Sciences} \\ {\em
 Novosibirsk,
 630090,
 Russia}
\\ {\em E-mail address: khatsym@gmail.com}}
\date{}
\maketitle

\begin{abstract}

This article is in the spirit of our work on the consequences of the Regge calculus, where some edge length scale arises as an optimal initial point of the perturbative expansion after functional integration over connection.

Now consider the perturbative expansion itself. To obtain an algorithmizable diagram technique, we consider the simplest periodic simplicial structure with a frozen part of the variables ("hypercubic"). After functional integration over connection, the system is described by the metric $g_{\lambda \mu}$ at the sites.

We parameterize $g_{\lambda \mu}$ so that the functional measure becomes Lebesgue. The discrete diagrams are free from ultraviolet divergences and reproduce (for ordinary, non-Planck external momenta) those continuum counterparts that are finite. We give the parametrization of $g_{\lambda \mu}$ up to terms, providing, in particular, additional three-graviton and two-graviton-two-matter vertices, which can give additional one-loop corrections to the Newtonian potential.

The edge length scale is $\sim \sqrt{ \eta }$, where $\eta$ defines the free factor $ ( - \det \| g_{\lambda \mu} \| )^{ \eta / 2}$ in the measure and should be a large parameter to ensure the true action after integration over connection. We verify the important fact that the perturbative expansion does not contain increasing powers of $\eta$ if its initial point is chosen close enough to the maximum point of the measure, thus justifying this choice.

Discrete propagators depend on the Barbero-Immirzi parameter $\gamma$, which determines the ratio of timelike and spacelike elementary length scales. The existing estimates of $\gamma$ allow the propagator poles to have real energy for any (real) spatial momenta.

\end{abstract}


MSC classes: 83C27; 81S40

keywords: general relativity; piecewise flat spacetime; Regge calculus; discrete connection; functional integral

\section{Introduction}

The form of coordinateless regularization of general relativity (GR) proposed by Regge in Ref. \cite{Regge}, Regge calculus (RC), can itself be considered as a minisuperspace theory, that is, the same GR, but on a subset in the configuration superspace. Each point of the configuration superspace is some Riemannian geometry, and the subset consists of the piecewise flat geometries. This subset is dense in the superspace in some topology, that is, an arbitrary Riemannian geometry can be approximated by piecewise flat geometries in some topology \cite{Fein,CMS}. A piecewise flat manifold is defined as having zero measure of curvature support and can be considered as glued together from flat 4-simplices (4-dimensional tetrahedra). Such a geometry is completely characterized by a discrete set of variables - the lengths of the edges of the tetrahedra. Discreteness is important when trying to quantize GR, the continuum nature of which leads to its formal non-renormalizability. Thus, RC proves to be effective in the path integral approach to gravity, including obtaining physical quantities such as the Newtonian potential \cite{HamWil1,HamWil2,Ham1}. To reproduce a smooth metric distribution on a large scale, a few fixed types of building microblocks (4-simplices) are sufficient. Restriction to them simplifies the calculation of cumbersome functional integrals. This approach is developed in the Causal Dynamical Triangulations theory (CDT) \cite{cdt,cdt1}. RC was used mainly as a regularization tool, but the approach of Ref. \cite{Mik} assumes that spacetime is indeed piecewise flat.

The above Newtonian potential seems to be a natural touchstone for applications of quantum Regge calculus \cite{HamWil1}. Along with this, the Newtonian potential was studied in the continuum perturbation theory. A general analysis of one-loop divergences in gravity is given in Ref. \cite{hooft}. One-loop quantum corrections to the Newtonian potential due to the exchange of gravitons in perturbation theory were calculated in a number of works \cite{don,don1,don2,don3,muz,akh,hamliu1,kk,kk1}. Naturally, the problem arises of calculating such Feynman diagrams in higher orders, which diverge and require taking into account the real simplicial structure of spacetime (as long as the Regge calculus is used for regularization), when calculating the Newtonian potential and other physical quantities. Or, in other words, the problem of constructing a diagram technique in simplicial gravity. Such a diagram technique is considered in Ref. \cite{HamLiu}. Of course, if one simply formally writes down a continuous theory on a lattice, then the Feynman diagrams which were finite in this theory, in the discrete version will be almost the same in magnitude for ordinary external momenta or for distances much larger than the typical edge length scale, which plays the role of a lattice spacing.

In our case, as a source of ultraviolet (UV) cutoff, which does not allow edge lengths to be arbitrarily small with any appreciable probability (and even loosely fixes edge lengths around a finite nonzero scale), we can consider some properties of the discrete functional integral measure \cite{our1}. This measure provides additional graviton vertices and additional diagrams to the discrete counterparts of the continuum diagrams and vertices in the discrete diagram technique.

If we briefly consider this in more detail, such a measure arises upon functional integration over the connection variables. Connection variables are introduced, because such attributes of a rigorous approach to constructing a functional integral, such as the continuous time limit and Hamilton's canonical formalism, are not well-defined in terms of exclusively edge length variables. But extending the set of variables by adding the connection variables introduced by Fr\"{o}hlich \cite{Fro} improves the situation. We take the RC action as a sum of contributions from self-dual and anti-self-dual parts of SO(3,1) rotations taken as independent connection variables \cite{Kha}. The remaining variables (of the tetrad type) are edge vectors in the local frames associated with individual 4-simplices. Excluding the connection via the equations of motion would give the genuine Regge action.

Attention here can be drawn to a possible parallel with Loop Quantum Gravity (LQG) in Ashtekar's variables \cite{Ash}, which are based on the use of the (anti-)self-dual su(2) connection. We use both connections, self-dual and anti-self-dual ones. In the continuum case, this would be equivalent to using the full connection (generalized Palatini) form of the Hilbert-Einstein action, but in the discrete case, due to the non-linearities of finite rotations, the decomposition of the total SO(3,1) into self-dual and anti-self-dual parts can lead to a quantum-nonequivalent (when connection is a dynamical variable), but technically a more manageable theory. In addition, LQG calculates the area spectrum, which turns out to be discrete with a quantum proportional to the Planck scale and the Barbero-Immirzi parameter $\gamma$ \cite{RovSmo}; we calculate the probability distribution of the elementary Regge areas or their vacuum averages, and this is a different characteristic than the area spectrum: the typical area scale (ordinary or spacelike) defined in this way is independent of $\gamma$ in the first approximation (whereas the timelike area scale is proportional to $\gamma$).

The aforementioned connection representation allows us to define the functional measure in the continuous time formalism and further address the problem of finding a complete discrete measure such that it becomes the found version for continuous time, no matter which direction is taken as the direction of time, with the geometry smoothed along it by setting the edge lengths arbitrarily small in this direction. This problem is resolvable (with slightly modified conditions: in the extended superspace of independent area tensors and with the operation of projecting onto the physical hypersurface in this superspace), and the full discrete measure can be found. It is defined up to a parameter $\eta$, whose change $\Delta \eta$ leads to additional factors in the measure $V^{\Delta \eta}_{\sigma^4}$ or $ ( - \det \| g_{\lambda \mu} \| )^{\Delta \eta / 2}$ in the continuum theory, where $V_{\sigma^4}$ is the 4-volume of the 4-simplices $\sigma^4$.

With this measure $\d \mu ( \rl ) {\cal D} \Omega$ on the tetrad type variables $\rl$ and the connection $\Omega \in$ SO(3,1) (${\cal D} \Omega$ is the Haar invariant measure) and the mentioned representation of the RC action $S_{\rm g}(\rl , \Omega )$, the functional integration over ${\cal D} \Omega$ can be performed leading to the resulting functional integral with a phase $\tilS_{\rm g} ( \rl )$ and a measure $F ( \rl ) D \rl$,
\begin{equation}\label{int-S D-Omega=F}                                     
\int \exp [ i S_{\rm g}(\rl , \Omega  ) ] ( \cdot ) \d \mu ( \rl ) {\cal D} \Omega = \int \exp [ i \tilS_{\rm g} ( \rl ) ] ( \cdot ) F ( \rl ) D \rl .
\end{equation}

\noindent It is appropriate to calculate the argument $\tilS_{\rm g} ( \rl )$ and modulus $F ( \rl ) D \rl$ of the result of this integration using such expansions of $S_{\rm g}(\rl , \Omega )$ that lead to the main contribution to the function of interest already in the zeroth order.

A nonzero contribution to the phase $\tilS_{\rm g} ( \rl )$ arises already in the zeroth order when expanding $S_{\rm g}(\rl , \Omega ) = \Omega_0 \exp \omega$ over $\omega \in$ so(3,1) in the vicinity of some solution to the equations of motion for $\Omega$, that is, in the vicinity of the background $\Omega = \Omega_0 ( \rl )$. Then $S_{\rm g}(\rl , \Omega ) - S_{\rm g}(\rl , \Omega_0 ( \rl ) )$ begins with a bilinear form over $\omega$. Then the higher orders in $\omega$ can be considered as corrections, and the Gaussian integral over $\omega$ gives a series in powers of $l_{\rm Pl} / l$, where $l$ is a typical edge length scale and $l_{\rm Pl}$ is the Planck length. $l_{\rm Pl} / l$ is a small parameter at $l \gg l_{\rm Pl}$. Looking ahead, in the considered approach, the edge length scale arises dynamically (as a characteristic of an optimal starting point of the perturbative expansion) as spacelike $b_{\rm s}$ and timelike $b_{\rm t}$ lattice steps, which depend on $\eta$ (eq (\ref{l=l-pl-sqrt-eta})); to ensure $l \gg l_{\rm Pl}$, $\eta$ should be a large parameter. By definition of the connection representation $S_{\rm g}(\rl , \Omega )$, the zeroth order term $S_{\rm g}(\rl , \Omega_0 ( \rl ) )$ in $\tilS_{\rm g} ( \rl )$ is just the RC action $S_{\rm g} ( \rl )$.

For $F ( \rl )$, the expansion over some edge vectors $l^a_\lambda$, which can be called {\it temporal}, discrete analogs of the tetrad $e^a_\lambda$ with the world index $\lambda = 0$ (the ADM lapse-shift functions \cite{ADM1} $(N, \bN )$), allows to get the main part of the contribution already in the zeroth order. In this order, we get the "factorization approximation", when the temporal triangles do not contribute to the action $S_{\rm g}(\rl , \Omega )$, and the contribution of other, spatial and diagonal triangles to the result of the functional integration over connection is factorized over individual triangles.

In the application considered in this article, we are faced with the need to have the dynamics (of discrete analogues) of all four tetrad vectors. Fortunately, due to the symmetry between different directions, we can analogously consider such an expansion in $l^a_\lambda$ with any other index $\lambda$ and get a similar factorization of the result of the functional integration over connection over triangles in zero order. And all this for different $\lambda$ can be reduced to a single formula with the same accuracy.

To develop the perturbative expansion for the functional integral (\ref{int-S D-Omega=F}), we should pass from $\rl = (l_1, \dots, l_n )$ to new variables $u = (u_1, \dots, u_n )$, in which the measure becomes Lebesgue, $F ( \rl ) D \rl = D u$. Then $\tilS_{\rm g} ( \rl )$ can be expanded in the vicinity of some initial point $\rl_{(0)} = \rl ( u_{(0)} )$,
\begin{equation}\label{ddS}                                                 
\tilS_{\rm g} ( \rl ) = \tilS_{\rm g} ( \rl_{(0)} ) + \frac{1}{2} \sum_{j, k, l, m} \frac{\partial^2 \tilS_{\rm g} ( \rl_{(0)} )}{\partial l_j \partial l_l} \frac{\partial l_j (u_{(0)} )}{\partial u_k} \frac{\partial l_l (u_{(0)} )}{\partial u_m} \Delta u_k \Delta  u_m + \dots .
\end{equation}

\noindent Here $\Delta u = u - u_{(0)}$. Regge's skeleton equations (if $\tilS_{\rm g} ( \rl ) = S_{\rm g} ( \rl )$ is taken) should also hold at this point,
\begin{equation}\label{dS/dl=0}                                             
\frac{\partial \tilS_{\rm g} ( \rl_{(0)} )}{\partial l_j} = 0 .
\end{equation}

\noindent Similar to imposing the extremum condition (\ref{dS/dl=0}) on the zeroth order term to maximize the contribution, for the same purpose it is possible to impose the minimization condition for the determinant of the second order form in the exponent,
\begin{equation}\label{def-l0}                                              
F (\rl_{(0)} )^{-2} \det \left \| \frac{\partial^2 \tilS_{\rm g} (\rl_{(0)} )}{\partial l_i \partial l_k} \right \| = \mbox{ minimum}.
\end{equation}

\noindent To be exact, here $\tilS_{\rm g}$ is replaced by $S$, which also includes gauge-breaking and Faddeev-Popov ghost terms; condition (\ref{def-l0}) is satisfied on the spacelike $ l = b_{\rm s} $ and timelike $ l = b_{\rm t} $ lengths (Subsection \ref{starting-point}), which thus form lattice steps, here in usual units, $\gamma$ is the Barbero-Immirzi parameter:
\begin{equation}\label{l=l-pl-sqrt-eta}                                     
b_{\rm s} = l_{\rm Pl} \sqrt{ \frac{ \eta - 10 }{ 3 \pi } }, ~~~ b_{\rm t} = \gamma b_{\rm s}, ~~~ l_{\rm Pl} = \sqrt{8 \pi G} .
\end{equation}

Thus, the maximum of the measure $F ( \rl ) D \rl $ allows us to loosely fix the edge length scales  $ b_{\rm s} $, $ b_{\rm t} $ as defining an optimal initial point for the perturbative expansion of the functional integral. (We can imagine an analogy with a body spontaneously choosing an equilibrium position in a potential field.) We can say that the scales $ b_{\rm s} $, $ b_{\rm t} $ are dynamically defined inside simplicial gravity itself. They are proportional to the Planck length and thus to $\sqrt{ \hbar }$. In the classical limit $\hbar \to 0$, we have $b_{\rm s} ,  b_{\rm t} \to 0$, that is, passing to continuum. Thus, discreteness here is a quantum effect.

Besides that, RC is a unique case when the field variables (the edge lengths) simultaneously play the role of steps of some lattice. Therefore, we can speak of a {\it dynamical lattice}, the spacings of which are loosely fixed within the theory itself.

Having got such a dynamical lattice, we can develop a perturbation theory on this lattice theory, and the main non-perturbative effect is already contained in the nonzero $ b_{\rm s} $, $ b_{\rm t} $, i.e., in the lattice itself; the residual non-perturbative effect is indirect: the same measure $F ( \rl ) D \rl $ that fixes $ b_{\rm s} $, $ b_{\rm t} $ affects the perturbative expansion by adding new vertices to it, and we analyze them here.

The (dynamical) lattice already serves as an UV regulator, which actually leads to a momenta cutoff at $ b_{\rm s}^{- 1} $, $ b_{\rm t}^{- 1} $; on lattice perturbative diagrams in the momentum representation, this is manifested in the fact that they are expressed as integrals over the components of quasi-momenta from finite regions ($[- \pi, \pi ]$), and not over momenta from the entire number axis.

The above loose fixation of edge lengths around some scale turns out to be useful in studying the very initial point of the perturbative expansion for the functional integral, when this point is considered to be an extreme background like a black hole. We have used this approach to resolve singularities in the main types of black holes and estimate the metric/field distribution in the vicinity of these former singularities \cite{our}. Now we consider a flat background.

For diagrams, we need propagators. To write the graviton propagator in a closed form, we take the simplest periodic simplicial structure with a hypercubic cell divided by diagonals into 4!=24 4-simplices \cite{RocWil} and freeze some of the tetrad-connection variables, but in such a way that the remaining freedom is still sufficient for approximation in some topology of any smooth Riemannian manifold with arbitrary accuracy. In this case, the variables $\rl$ have the meaning of the metric $g_{\lambda \mu}$ per site/hypercube, and $\rl = \rl ( u )$ is a relation between $g_{\lambda \mu}$ and $u$, to which we will give the meaning of some new metric $\tg_{\lambda \mu}$.

The RC strategy implies taking into account all simplicial structures. This, in particular, would restore the symmetry broken by the choice of lattice axes. The diffeomorphism symmetry is preserved in the leading order over metric variations from 4-simplex to 4-simplex (or from site to site) in configurations in which such metric variations are small (that is, in configurations close to the continuum spacetime). This means the appearance of sets of infinite measure of physically almost equivalent configurations. Since the diffeomorphism symmetry is still broken by discreteness, we cannot simply fix the gauge to eliminate multiple counting by introducing a delta function, as this would mean canceling some physical degrees of freedom. Therefore, for the discrete case, it is more expedient not to fix the gauge conditions, but to average over some neighborhood of such conditions, introducing a term whose gauge non-invariance manifests itself here in the leading order in metric variations. In fact, such a relevant term turns out to be close to the usual continuum term corresponding to averaging in the vicinity of the de Donder-Fock gauge.

As mentioned above, $\eta$ should be a large parameter. It is not a priori obvious that the perturbative expansion will not contain increasing powers of $\eta$. We will see that this is indeed the case, but only if the initial point of the perturbative expansion is chosen close enough to the maximum point of the measure.

And this can serve as a dynamic substantiation of the choice of the initial point of the expansion in the vicinity of the maximum point of the measure.

If continuum diagrams converge (for example, one-loop corrections to the Newtonian potential), their discrete versions are almost the same in magnitude (up to lattice corrections, which are small for ordinary momenta). Of course, for UV divergent continuum diagrams, their discrete versions remain free of UV divergences.

We have found $g_{\lambda \mu} = g_{\lambda \mu} ( \{ \tg_{\nu \rho} \} )$ up to bilinear terms over (variations of) $\tg_{\lambda \mu}$. These terms lead, in particular, to additional three-graviton and two-graviton-two-matter vertices, which give one-loop corrections to the Newtonian potential.

The above $\rl = \rl ( u )$ (or, in particular, $g_{\lambda \mu} = g_{\lambda \mu} ( \{ \tg_{\nu \rho} \} )$) is not, generally speaking, unique. Our "physically" (or rather geometrically) natural choice is based on factorization of the measure in the variables $\rl$, chosen so that some of them have the meaning of some areas $\rv_1 , \dots , \rv_m$: $F ( \rl ) D \rl = [\prod^m_{k = 1} f_k ( \rv_k ) \d \rv_k ] \prod^n_{k = m + 1} \d l_k$, and $D u$ can be written in $\trv_1 , \dots , \trv_m , \tl_{m + 1} , \dots , \tl_n$, which are the prototypes of $\rv_1 , \dots , \rv_m , l_{m + 1} , \dots , l_n$ for the metric $\tg_{\lambda \mu}$: $D u = [\prod^m_{k = 1} \tf_k ( \trv_k ) \d \trv_k ] \prod^n_{k = m + 1} \d \tl_k$. (This is up to perturbative corrections, which then can be calculated.) Then it is natural to set $f_k ( \rv_k ) \d \rv_k = \tf_k ( \trv_k ) \d \trv_k$, $k = 1, \dots , m$, $l_k = \tl_k$, $k = m + 1 , \dots , n$.

We also consider possible initial stages of evaluating diagrams. In contrast to the case of a continuum, depending on the ratio of timelike and spacelike elementary length scales $b_{\rm t} / b_{\rm s} = \gamma $, a discrete propagator may not have a pole at real energy for some (real) spatial momenta. As we will consider further (Subsection \ref{typical}), the existing estimates of $\gamma$ satisfy the condition (\ref{gamma<...}) required for the poles to have real energy for any spatial momenta.

The measure has the same form in $h^{\lambda \mu} = g^{\lambda \mu} \sqrt{ - g }$ as in $g_{\lambda \mu}$, and this method admits an alternative {\it dual} construction giving the dual parametrization $h^{\lambda \mu} = h^{\lambda \mu} ( \{ \tih^{\nu \rho} \} )$. This possibility is considered.

The paper is organized as follows. In Section \ref{starting}, the starting functional integral in terms of metric/tetrad/lengths variables is specified, in particular, the action is considered in Subsection \ref{action}, and the measure - in Subsection \ref{measure}. The form of a term whose gauge non-invariance manifests itself in the leading order in metric variations is considered in Subsection \ref{gauge}; in fact, it is similar to the usual continuum term, corresponding to averaging in the vicinity of the de Donder-Fock gauge, and is tuned to meet the needs of the problem at hand (to be proportional to the scale [length]$^2$). The corresponding Faddeev-Popov ghost contribution is considered in Subsection \ref{ghost}. In these Subsections \ref{gauge}, \ref{ghost}, we use some notation from the book \cite{KonPop}. In Section \ref{reduction}, the functional integral is reduced to a form appropriate for constructing perturbation series. In Subsection \ref{splitting}, the problem of transforming the functional measure into a Lebesgue one is divided, with some corrections, into one-dimensional problems. In Subsection \ref{starting-point}, the optimal initial point of the perturbative expansion is found; it includes different timelike and spacelike scales. Subsection \ref{one-dimensional} considers the obtained one-dimensional problem and analyzes its solution in the form of an expansion. In Subsection \ref{large-parameter}, this expansion leads to the absence of a large parameter in the perturbative expansion. In Subsection \ref{correction} we consider a typical correction to the splitting approximation and find that it does not lead to a large parameter in the perturbative expansion. Subsection \ref{maximum} shows that a large parameter appears in the perturbative expansion if the starting point is taken far enough from the maximum point of the measure. In Subsection \ref{parametrization}, we find the parametrization of the metric $g_{\lambda \mu} = g_{\lambda \mu} ( \{ \tg_{\nu \rho} \} )$ up to bilinear terms. These terms lead to additional three-graviton, two-graviton-two-matter and two-graviton-two-ghost vertices. In Subsection \ref{propagators}, the graviton and ghost propagators are written out in the quasi-momentum representation. In Subsection \ref{general}, we consider some features of the general expansion parameterizing the metric, typical order in $\eta$ of a multigraviton and multigraviton-two-ghost vertex, and reproducing converging continuum diagrams from their discrete counterparts. In Subsection \ref{typical}, this transition of the discrete to continuum diagrams is illustrated by examples of some typical diagrams, and possible initial stages of evaluating diagrams are considered. In Subsection \ref{dual}, we consider the dual construction of the metric parametrization, $h^{\lambda \mu} = h^{\lambda \mu} ( \{ \tih^{\nu \rho} \} )$, together with the graviton propagator. Then comes the Conclusion.

\section{Starting functional integral}\label{starting}

\subsection{The action}\label{action}

The original action $S_{\rm g}(\rl , \Omega )$ or, more accurately, that one in terms of area tensors $v$ (or complex 3-vectors $\bv$) and $\Omega$ is

\begin{eqnarray}\label{S-simplicial}                                        
S_{\rm g} [ v, \Omega ] = \frac{1}{2} \sum_{\sigma^2} \left ( 1 + \frac{i}{\gamma } \right ) \sqrt{ \bv^2_{\sigma^2} } \arcsin \frac{\bv_{\sigma^2} * \pR_{\sigma^2} ( \Omega )}{\sqrt{ \bv^2_{\sigma^2}}} + {\rm c. c. }
\end{eqnarray}

\noindent (c.c. = complex conjugate). The connection matrices $\Omega_{\sigma^3}$ "live" on the tetrahedra $\sigma^3$ (3-simplices), while on the triangles $\sigma^2$, the holonomy of $\Omega$ or the curvature $R_{\sigma^2} ( \Omega ) = \prod_{{\sigma^3} \supset {\sigma^2}} \Omega^{\pm 1}_{\sigma^3}$ is defined. The decomposition of an SO(3,1) matrix into self-dual and anti-self-dual ones is used, $\Omega = \pOmega \mOmega , \mOmega = ( \pOmega )^* , \pR ( \Omega ) = R ( \pOmega )$, according to the decomposition of SO(3,1) as (a subgroup of) SO(3,C) $\times$ SO(3,C). If a triangle $\sigma^2$ is formed by two edges $\sigma^1_1 $, $\sigma^1_2$, then $2 \bv_{\sigma^2} = i \bl_{\sigma^1_1} \times \bl_{\sigma^1_2} - \bl_{\sigma^1_1} l^0_{\sigma^1_2} + \bl_{\sigma^1_2} l^0_{\sigma^1_1}$. $\pmOmega$ and $\pmR$ are represented as 3 $\times$ 3 matrices, and $\bv * R \equiv \frac{1}{2}v^i R^{kl} \epsilon_{ikl}$.

The simplest periodic simplicial structure consists of hypercubes, each of which is divided by diagonals into 24 4-simplices. The action (\ref{S-simplicial}) can be written for this structure. We are going to further simplify this system by giving specific values to some area tensors and connection matrices. Namely,

1) $\Omega_{\sigma^3} = \Omega_\lambda$ for each of the six $\sigma^3$ into which the 3D face (the 3-cube, orthogonal to the edge along $x^\lambda$) is divided,

2) $\Omega_{\sigma^3} = 1$ for the internal $\sigma^3$s in the 4-cube,

3) the area tensors of the two triangles that make up the quadrilateral in the plane of $x^\lambda, x^\mu$ are the same.

\noindent The resulting "hypercubic" action takes the form
\begin{eqnarray}\label{S-cube}                                              
S_{\rm g} [ v, \Omega ] = \frac{1 }{4 } \sum_{\stackrel{\scriptstyle \lambda \mu \nu \rho}{\rm sites}} \left ( 1 + \frac{i}{\gamma } \right ) \epsilon^{\lambda \mu \nu \rho} \sqrt{ \bv^2_{\lambda \mu} } \arcsin \frac{\bv_{\lambda \mu} * \pR_{\nu \rho} ( \Omega )}{\sqrt{ \bv^2_{\lambda \mu}}} + {\rm c. c. } .
\end{eqnarray}

\noindent where
\begin{equation}\label{v=ll-R=wwww}                                         
2 \bv_{\lambda\mu} = i \bl_\lambda \times \bl_\mu + l^0_\lambda \bl_\mu - l^0_\mu \bl_\lambda, ~~~ R_{\lambda\mu} (\Omega ) = \overOm_\lambda (\overT_\lambda \overOm_\mu) (\overT_\mu \Omega_\lambda) \Omega_\mu .
\end{equation}

\noindent Here $T_\lambda$ is the shift operator acting as $T_\lambda f(\dots , x^\lambda , \dots ) = f(\dots , x^\lambda + 1 , \dots )$ for a function $f$; $\overline{T}_\lambda$, $\overline{\Omega}_\lambda$ are the Hermitian conjugates of the operator $T_\lambda$ and matrix $\Omega_\lambda$, which coincide with $T^{-1}_\lambda$ and $\Omega^{-1}_\lambda$, respectively. $\epsilon^{0123} = +1$.

Excluding $\Omega$ from the action $S_{\rm g} [ v, \Omega ]$ with the help of the equations of motion, we get an action as a function of the metric $ \tilS_{\rm g} [ g ] $. In \cite{our1}, we have found that, in the leading order in variations between sites of the vectors $l^a_\lambda$, the procedure of excluding $\Omega$ from (\ref{S-cube}) reduces to the procedure of excluding $\omega \in so(3,1)$ ($\Omega_\lambda \equiv \exp \omega_\lambda$) from
\begin{equation}\label{v-D-omega}                                                
- \frac{1}{4} \sum_{\rm sites} \left ( 1 + \frac{i}{\gamma } \right ) \epsilon^{\lambda \mu \nu \rho} \bv_{\lambda \mu} \cdot (\Delta_\nu \bomega_\rho - \Delta_\rho \bomega_\nu + \bomega_\nu \times \bomega_\rho) + {\rm c. c. }.
\end{equation}

\noindent $\omega^k_\lambda = - \frac{1}{4} \epsilon^k{}_{l m} \omega^{l m}_\lambda + \frac{i}{2} \omega^{0 k}_\lambda$ are components of the appearing here three-dimensional vector $\bomega_\lambda$ (in superscript). $\epsilon_{123} = +1$. $\Delta_\lambda = T_\lambda - 1 $ is a finite difference operator. Eq. (\ref{v-D-omega}) is some finite-difference form of the continuum tetrad-connection representation \cite{Holst,Fat} of the Einstein action $\frac{1}{2} \int R \sqrt{-g} \d^4 x$,
\begin{eqnarray}\label{Cartan}                                             
& & S_{\rm g} [ e, \omega ] = - \frac{1}{8}\int{(\epsilon_{abcd}e^a_{\lambda}e^b_{\mu} + \frac{2}{ \gamma}e_{\lambda c}e_{\mu d })\epsilon^{\lambda\mu\nu\rho} [\partial_{\nu} + \omega_{\nu}, \partial_{\rho} + \omega_{\rho}]^{cd}{\rm d}^4x} = - \frac{1}{8} \epsilon^{\lambda\mu\nu\rho} \cdot \nonumber \\ & & \hspace{-2mm} \cdot \left ( 1 + \frac{i}{\gamma } \right ) \int (i \be_\lambda \times \be_\mu + e^0_\lambda \be_\mu - e^0_\mu \be_\lambda) \cdot (\partial_\nu \bomega_\rho - \partial_\rho \bomega_\nu + \bomega_\nu \times \bomega_\rho) \d^4 x + {\rm c. c. },
\end{eqnarray}

\noindent with the tetrad $e^a_\lambda = l^a_\lambda$ and the coordinate differences $\Delta x^\lambda = 1$ between the neighboring sites. In the leading order, finite differences obey the same rules as ordinary derivatives, and a finite-difference form of the tetrad action, as a result of the elimination of $\omega$, is reduced to a finite-difference form of the Einstein action $\frac{1}{2} \int R \sqrt{-g} \d^4 x$,
\begin{eqnarray}\label{Sg[g]}                                               
\tilS_{\rm g} [ g ] & = & \frac{1}{ 8 } \sum_{\rm sites} ( \Delta_\nu g_{\lambda \mu} ) ( \Delta_\tau g_{\rho \sigma} ) ( 2 g^{\lambda \rho} g^{\mu \tau} g^{\nu \sigma} - g^{\lambda \rho} g^{\mu \sigma} g^{\nu \tau} - 2 g^{\lambda \tau} g^{\mu \nu} g^{\rho \sigma} \nonumber \\ & & \phantom{- \frac{1}{ 64 \pi G } \int ( \Delta_\nu g_{\lambda \mu} ) ( \Delta_\tau g_{\rho \sigma} ) (} + g^{\lambda \mu} g^{\rho \sigma} g^{\nu \tau} ) \sqrt{ - g }
\end{eqnarray}

\noindent with the metric $g_{\lambda \mu} = \eta_{a b} l^a_\lambda l^b_\mu$. Of course, we cannot decompose a curved piecewise flat spacetime using hypercubes so that the edge vectors are completely described by a tetrad at each vertex. But we are mathematically modeling a real piecewise flat spacetime, for which the exclusion of the connection from representation (\ref{S-simplicial}) leads to the exact Regge action. In terms of the contravariant density $h^{\lambda \mu} = g^{\lambda \mu} \sqrt{ - g } $, the action has the form
\begin{eqnarray}\label{Sg[g](h)}                                           
\tilS_{\rm g} [ g ] & = & \frac{1}{ 8 } \sum_{\rm sites} [ - 2 h^{\lambda \nu} ( \Delta_\lambda h^{\mu \rho} ) ( \Delta_\rho h_{\mu \nu} ) + h^{\lambda \rho} ( \Delta_\lambda h^{\mu \nu} ) ( \Delta_\rho h_{\mu \nu} ) \nonumber \\ & & \phantom{- \frac{1}{ 64 \pi G } \int (} + \frac{1}{2} h^{\lambda \mu} ( \Delta_\lambda \ln h ) ( \Delta_\mu \ln h )] .
\end{eqnarray}

\subsection{The measure}\label{measure}

In a synchronous gauge ($(N, \bN ) = (1, {\bf 0})$ for discrete analogs of the ADM lapse-shift functions \cite{ADM1} $(N, \bN )$ (certain edge vectors)), i.e., when trying to fit the transverse graviton wave to three dimensions, we got \cite{our1} a singularity at long times, $\sin^{-2} \! \! p_0$ in the graviton propagator. Thus, we should allow variations of the discrete $(N, \bN )$ as field variables.

That is, the functional integration in the tetrad/length sector is carried out over the full metric tensor at each site. The measure of such an integration, according to our assumption that it is not necessarily a scalar, but a scalar density, is equal to $\d^{10} g_{\lambda \mu}$ times a power of $-g = - \det \| g_{\lambda \mu} \|$ at each site. Our notation for this power stems from the procedure of projecting the measure from an extended configuration superspace of independent area tensors $v^{ab}_{\lambda\mu}$ onto the physical hypersurface where the conditions $\epsilon_{abcd} v^{ab}_{\lambda\mu} v^{cd}_{\nu\rho} \sim \epsilon_{\lambda\mu\nu\rho}$ are enforced with the help of the $\delta$-function factor $\delta_{\rm metric}$,
\begin{equation}\label{delta(vv)}                                          
\delta_{\rm metric} ( v ) = \int V^\eta \delta^{21} \left ( \epsilon_{abcd}v^{ab}_{\lambda\mu}v^{cd}_{\nu\rho}
- V \epsilon_{\lambda\mu\nu\rho} \right ) \d V ,
\end{equation}

\noindent ensuring that the area tensors correspond to certain edge vectors in each 4-simplex (here in each 4-cube). The product of such expressions over the 4-cubes or sites should be taken. In each 4-cube, the measure on the physical hypersurface is
\begin{equation}\label{int-d36v-delta-metric}                              
\int \d^{36} v^{a b}_{\lambda \mu} \delta_{\rm metric} ( v ) \sim ( - g )^\frac{ \eta - 7 }{ 2 } \d^{10} g_{\lambda \mu} .
\end{equation}

Besides that, the dependence on the discrete lapse-shift functions is also present in additional measure factors that arise as a result of the functional integration over connection. When calculating in Ref. \cite{Kha3}, we considered the case when we neglect the contribution of the discrete lapse-shift functions in the action, and then the result of the integration over connection is factorized over spatial areas. In the case of our hypercubic model, there are three spatial areas per site with the area vectors $\bv_\alpha \equiv \epsilon_\alpha{}^{\beta \gamma} \bv_{\beta \gamma} / 2$ ($\alpha , \beta , \gamma , ... = 1, 2, 3$), and the measure is
\begin{equation}\label{N0-v-alpha}                                         
\prod_{\rm sites} \left ( \prod_\alpha \N_0 (2 \rv_\alpha ) \right ) ( - g )^\frac{ \eta - 7 }{ 2 } \d^{10} g_{\lambda \mu}.
\end{equation}

\noindent Here $\rv = \sqrt{\bv^2}$, the complex area vectors $\bv_{\lambda\mu}$ are expressed in terms of edge vectors (constituting a tetrad $l^a_\lambda$ in our model) according to equation (\ref{v=ll-R=wwww}), and
\begin{equation}\label{N0}                                                 
\N_0 (\rv ) = \left | \frac{1}{\frac{1}{4} \left ( \frac{1}{\gamma} - i \right )^2 \rv^2 + 1} \frac{\frac{1}{4} \left ( \frac{1}{\gamma} - i \right ) \rv}{ \sh \left [ \frac{\pi}{2} \left ( \frac{1}{\gamma} - i \right ) \rv \right ]} \right |^2
\end{equation}

\noindent is the aforementioned factor due to the functional integration over connection \cite{Kha3}. As it stands with the argument $\rv$, it arises when working with the general Regge manifold for a triangle with the area $\rv$; in the hypercubic model, triangles appear in pairs of identical ones, which determines the argument $2 \rv_\alpha$ of $\N_0$ in (\ref{N0-v-alpha}).

The terms "temporal" and "spatial" refer to the discrete analogs of the world vector subscript $\lambda = 0$ and $\lambda = 1, 2, 3$, respectively, and mean certain edges; as applied to triangles, these mean the triangles containing a temporal edge or not, respectively. The terms "timelike" and "spacelike" refer to the discrete analogs of the local vector index $a = 0$ and $a = 1, 2, 3$, respectively.

Analogously to the temporal direction, we can consider edge vectors in the direction of some other coordinate $x^\alpha$ and the case when we can neglect the contribution of the triangles containing these edges in the action. Then the result of the integration over connection is factorized over the triangles located in the $x^0 x^\beta x^\gamma$ hyperplane ($ \alpha \neq \beta \neq \gamma \neq \alpha $), and the measure looks like
\begin{equation}\label{N0-v-0beta-0gamma-alpha}                            
\prod_{\rm sites} \N_0 (2 \rv_{0 \beta} ) \N_0 (2 \rv_{0 \gamma} ) \N_0 (2 \rv_\alpha ) ( - g )^\frac{ \eta - 7 }{ 2 } \d^{10} g_{\lambda \mu}.
\end{equation}

In the general case, there is no direction, the contribution of the triangles with edges along which can be neglected. Then we have non-factorizable integrations over the full set of independent curvature matrices, whichever such a complete set is chosen, and the result can hardly be obtained in a closed analytic form. One could use the main term approximation of the expansion of the result over the discrete lapse-shift functions (\ref{N0-v-alpha}) as the initial one. However, at $\gamma^{-1} \neq 0 $ suppression of large timelike areas of the temporal triangles $\rv_{0 \alpha}$, $\rv_{0 \alpha}^2 > 0$, takes place as well, as is seen from the measure in the aforementioned regions admitting exact results like (\ref{N0-v-0beta-0gamma-alpha}). To take this into account (and simultaneously make the formula symmetric over the coordinate directions), we use some simplified model measure. This model measure should be reduced to (\ref{N0-v-alpha}) or (\ref{N0-v-0beta-0gamma-alpha}) when we neglect the contribution of certain sets of triangles by tending $\rv_{0 \alpha} \to 0$, $\alpha = 1, 2, 3$ or $\rv_\beta \to 0$, $\rv_\gamma \to 0$, $\rv_{0 \alpha} \to 0$, respectively. The simplest such expression is one that includes the factors on the triangles of all six types,
\begin{equation}\label{N0-v-alpha-0alpha}                                  
\prod_{\rm sites} \left ( \prod_\alpha \N_0 (2 \rv_\alpha ) \N_0 (2 \rv_{0 \alpha} ) \right ) ( - g )^\frac{ \eta - 7 }{ 2 } \d^{10} g_{\lambda \mu}.
\end{equation}

\noindent Note that we have finite nonzero $\N_0 ( 0 ) = (2 \pi )^{-2}$. But no less important is that the expansion of $\N_0 ( \rv ) $ goes over $\rv^2$: $\N_0 ( \rv ) \propto 1 + {\rm const} \cdot \rv^2 + \dots $. Thus, if we want to increase the fineness of the subdivision into simplexes along a certain coordinate by $n$ times and divide each quadrilateral into $n$ quadrangles, then $\N_0 (2 \rv )$ will be replaced by $[\N_0 (2 \rv / n )]^n \propto 1 + O ( 1 / n ) \to 1 $ at $n \to \infty$. The appearing here areas have different $i$-ness (normally $\rv_{0 \alpha}^2 > 0$, $\rv_\alpha^2 < 0$), but it is remarkable that due to the Barbero-Immirzi parameter $\gamma < \infty$ we have exponential suppression in both spacelike and timelike regions.

The meaning of taking into account the suppression of large areas of temporal triangles, proposed by the introduction of formula (\ref{N0-v-alpha-0alpha}), is to avoid the occurrence of a large parameter in the perturbative expansion considered in Subsection \ref{maximum} around equation (\ref{y=x^(1/(k+1))}).

\subsection{Gauge non-invariant term}\label{gauge}

It is understood that the gauge non-invariance manifests itself here in the leading order over metric variations.

The functional integral with the above measure will provide a loose fixation of the edge lengths on some scale. This means the formation of a dynamic lattice. But when describing gravitons propagating along this lattice using a metric, we encounter degrees of freedom close to the gauge degrees of freedom of the continuum theory, when the field/metric variations from 4-simplex to 4-simplex are small.

Due to the absence of exact symmetries in the discrete theory, the imposition of some conditions, taken as gauge ones, in the form of delta functions in the functional integral means the cancellation of some physical degrees of freedom. Therefore, the elimination of a set of physically almost equivalent configurations of infinite measure from the functional integral without rigidly fixing any conditions on these configurations, but by averaging over some neighborhood of such conditions, fits the discrete theory in the best possible way. So we have {\bf gauge averaging} rather than gauge fixing.

In the continuum theory, the action is invariant under the group of diffeomorphisms or coordinate transformations $\Xi$ generated by the infinitesimal coordinate transformations $\delta x^\lambda = \xi^\lambda ( x )$, in terms of which the transformations of the metric $\delta g_{\lambda \mu}$ are expressed by means of a differential operator,
\begin{equation}                                                           
(g_{\lambda \mu})^\Xi - g_{\lambda \mu} = \delta g_{\lambda \mu} = - g_{\lambda \nu } \partial_\mu \xi^\nu - g_{\mu \nu } \partial_\lambda \xi^\nu - \xi^\nu \partial_\nu g_{\lambda \mu } .
\end{equation}

In the connection representation of the genuine Regge calculus, we have exactly the Einstein-Regge action when the connection is excluded, and in the leading order over metric variations, we have a finite-difference form of the continuum Einstein action \cite{our2}. In non-leading orders over metric variations, this action is not described solely in terms of one tensor $g_{\lambda \mu}$ at each vertex, and the metric would have to be characterized by lattice-specific simplicial components additional to $g_{\lambda \mu}$. In the hypercubic model under consideration, we have a finite-difference form of the continuum Einstein action when the connection is excluded, also only in the leading order over metric variations. Therefore, $\delta g_{\lambda \mu}$ can be taken also in the leading order over metric variations, taking into account non-leading orders will be an excess of accuracy. In this order, the transformations of the metric look as
\begin{equation}\label{delta-g=xi}                                         
(g_{\lambda \mu})^\Xi - g_{\lambda \mu} = \delta g_{\lambda \mu} = - g_{\lambda \nu } \Delta_\mu \xi^\nu - g_{\mu \nu } \Delta_\lambda \xi^\nu - \xi^\nu \Delta_\nu g_{\lambda \mu } ,
\end{equation}

\noindent and we can talk about the invariance of the action with respect to these transformations in the leading order over metric variations from 4-simplex to 4-simplex. And if the continuum diagrams converge (an example is the one-loop corrections to the Newtonian potential), then in the discrete theory this corresponds to small quasi-momenta. Then we can neglect non-leading orders over metric variations from site to site. In cases where the contribution of large quasi-momenta is significant, the corresponding diagrams can be used for rough estimates of the considered effect.

To cancel infinite integration over (approximate) gauge group $\Xi$ in the functional integral $J ( \cdot ) = \int \exp ( i \tilS_{\rm g} [ g ] ) ( \cdot ) \d \mu [ g ]$ (here it is written as a functional on functionals of the metric to be averaged), we introduce a weight functional factor, which we write as $\exp ( i \cF [ g ] )$, together with the corresponding normalization factor $\Phi [ g ]$, thus inducing a linear mapping of the functionals,
\begin{equation}\label{J()->J(()PhiExp(iF)}                                
J ( \cdot ) \rightarrow J ( ( \cdot ) \Phi [ g ] \exp ( i \cF [ g ] ) ),
\end{equation}

\noindent where $\Phi [ g ]$ is defined by integrating $\exp ( i \cF [ g ] )$ over $\Xi$,
\begin{equation}\label{Phi*int-exp-F=1}                                    
\Phi [ g ] \int \exp{( i \cF [ g^\Xi ] )} \prod_{\rm sites} \d \Xi = 1 , ~~~ \d \Xi = \prod_\lambda \d \xi^\lambda .
\end{equation}

\noindent The factor $\exp ( i \cF [ g ] )$ should be non-invariant to ensure the convergence of the integral in (\ref{Phi*int-exp-F=1}). If the functional integral $J ( \cdot )$ were invariant, then mapping (\ref{J()->J(()PhiExp(iF)}) would not depend on $\cF [ g ]$ by construction.

However, introducing a particular lattice makes the properties of the measure (after the functional integration over connection leading to the aforementioned $\N_0 (2 \rv_{\lambda \mu} )$) singled out with respect to the lattice axes. The violation of the (approximate) gauge symmetry of the functional integral means that the result of the mapping of functionals (\ref{J()->J(()PhiExp(iF)}) applied to the considered functional integral (by introducing the factor $\Phi [ g ] \exp ( i \cF [ g ] )$ under the integral sign) will depend on the weight functional factor $\exp ( i \cF [ g ] )$. We can continue to use this approach, keeping in mind that adding functional-integral contributions from other simplicial structures with the same $\cF [ g ]$ and the same target symmetry transformation (\ref{delta-g=xi}) to the result will restore symmetry and independence from $\cF [ g ]$.

In the case of the Newtonian potential, the simplest such summation would be averaging over the orientations of the lattice with respect to the gravitating masses or, equivalently, averaging over the orientations of the vector connecting the masses with respect to the lattice axes.

Let us discuss a possible choice of $ \cF [ g ] $. We allow arbitrary constant metrics, so $ \cF [ g ] $ must contain finite differences, and they are derivatives in the continuous limit. Gaussian averaging seems to be a natural choice, since in the presence of derivatives it is the Gaussian functional integral that has an exact definition in the continuum; moreover, convergence is guaranteed for a wide set of averaged functionals. Thus, $ \cF [ g ] $ is a bilinear form of the metric variables $g$ containing finite differences. A natural requirement for $\cF$ is its Lorentz-invariant form in the continuum limit (not necessarily locally invariant). Also, $\exp ( i \cF )$ should have a stabilizing (cut-off) effect on four metric functions $\tf^\lambda$, so we could write something like $\cF \sim \tf^\lambda \eta_{\lambda \mu} \tf^\mu$. More precisely, we take $h^{\lambda \mu} = g^{\lambda \mu} \sqrt{ - g } $ as a natural contravariant metric variable convenient for writing the Einstein action and, say, the interaction with a scalar field, and set $\tf^\lambda$ such that $\tf^\lambda = \partial_\mu h^{\lambda \mu}$ in the continuum limit (then in this limit $\tf^\lambda = 0$ is the de Donder-Fock gauge). As a result, we write $\cF$ in the form
\begin{equation}\label{alpha-term}                                         
\cF [ g ] = - \frac{\alpha }{4 } \sum_{\rm sites} (\Delta_\nu h^{\lambda \nu}) ( - h )^{ - \frac{1}{4}} \teta_{\lambda \mu } ( \Delta_\rho h^{\mu \rho}) \equiv - \frac{\alpha }{4 } \sum_{\rm sites} f^\lambda [ g ] \teta_{\lambda \mu } f^\mu [ g ] .
\end{equation}

\noindent Here $f^\lambda [ g ] = (- h)^{- 1 / 8} \Delta_\mu h^{\lambda \mu}$, and $( - h )^{ - \frac{1}{4}} \teta_{\lambda \mu }$ ($h = \det \| h^{\lambda \mu} \| = g$) instead of the commonly used $\eta_{\lambda \mu } = {\rm diag}(-1,1,1,1)$ when fixing this gauge in the continuum GR is introduced in order to preserve the scaling property of the action (like $g_{\lambda \mu}$ or $h^{\lambda \mu}$ or [length]$^2$) when $\cF$ is added to it. This will simplify the estimation of the typical edge length scale using the maximization condition (\ref{def-l0}) including $\det \| \partial^2 S / \partial l_i \partial l_k \| $. Some constant $\teta_{\lambda \mu }$ (specified below by (\ref{tilde-eta})) is a rescaled $\eta_{\lambda \mu }$ in a certain way, taking into account some difference between the (dynamically fixed) timelike and spacelike scales, so that in the final transition to some effective metric field variable we would have ${\rm diag}(-1,1,1,1)$ for this in new axes and familiar expressions for propagators.

\subsection{Ghost contribution}\label{ghost}

It is more convenient not to directly evaluate $\Phi [ g ]$ from its definition (\ref{Phi*int-exp-F=1}), but perform averaging over $\kappa^\lambda$ with the exponential weight $\exp \left ( - i \frac{ \alpha }{ 4 } \sum_{\rm sites} \kappa^\lambda \teta_{\lambda \mu} \kappa^\mu \right )$ of the functional integral with the non-invariant delta-function factor $\prod_{\rm sites , \lambda} \delta (f^\lambda [ g ] - \kappa^\lambda )$ together with the corresponding normalization factor $\Phi_0 [ g ]$ inserted. The appearance of the delta-function factor under the functional integral sign means that we still impose some conditions on the lattice, thus cancelling some physical degrees of freedom due to the absence of any symmetry in general. However, this is only intermediate formal mathematical action, and subsequent integration when averaging over $\kappa^\lambda$ restores the cancelled physical degrees of freedom on the lattice, if any. Concerning this functional integral with the factor $\Phi_0 [ g ] \prod_{\rm sites , \lambda} \delta (f^\lambda [ g ] - \kappa^\lambda )$ inserted, we can repeat the discussion of the previous Subsection that if we can confine ourselves to the leading order over metric variations, then adding functional-integral contributions from other simplicial structures will restore symmetry and independence of the functional integral from the non-invariant factor, here from $\kappa^\lambda$. The independence of the functional integral from $\kappa^\lambda$ means that its exponential averaging leaves this integral the same up to an inessential constant. On the other hand, $\Phi_0 [ g ]$ is defined by the determinant of the operator ${\cal O}$, defined by the formula
\begin{equation}\label{22}                                                 
f^\lambda [ g ]^\Xi - f^\lambda = ((- h)^{- 1 / 8} \Delta_\mu h^{\lambda \mu})^\Xi - (- h)^{- 1 / 8} \Delta_\mu h^{\lambda \mu} = ({\cal O} \xi )^\lambda ,
\end{equation}

\noindent as

\begin{eqnarray}\label{23}                                                 
& & \Phi_0 [ g ] \prod_{\rm sites , \lambda} \delta (f^\lambda [ g ] - \kappa^\lambda ) \nonumber \\ & & = \left \{ \int \left[ \prod_{\rm sites , \mu} \delta (f^\mu [ g ]^\Xi - \kappa^\mu ) \right] \prod_{\rm sites} \d \Xi \right \}^{ - 1 } \prod_{\rm sites , \lambda} \delta (f^\lambda [ g ] - \kappa^\lambda ) \nonumber \\ & & = \left \{ \int \left[ \prod_{\rm sites , \mu} \delta [ ( {\cal O} \xi )^\mu ] \right] \prod_{\rm sites , \nu } \d \xi^\nu \right \}^{ - 1 } \prod_{\rm sites , \lambda} \delta (f^\lambda [ g ] - \kappa^\lambda ) \nonumber \\ & & = \det {\cal O} \prod_{\rm sites , \lambda} \delta (f^\lambda [ g ] - \kappa^\lambda ) ,
\end{eqnarray}

\noindent and the exponential averaging of the inserted factor $\Phi_0 [ g ] \prod_{\rm sites , \lambda} \delta (f^\lambda [ g ] - \kappa^\lambda )$ under the functional integral sign just gives $\exp ( i \cF [ g ] ) \det {\cal O}$. Thus, we effectively arrive at
\begin{equation}\label{24}                                                 
\Phi [ g ] = \det {\cal O} .
\end{equation}

\noindent It is convenient to write ${\cal O} = ( - h )^{ - 1 / 8} \tilde{\cal O}$, where
\begin{equation}\label{tilde-O}                                            
(\tilde{\cal O} \xi )^\lambda = \Delta_\mu [ h^{\mu \nu} \Delta_\nu \xi^\lambda - ( \Delta_\nu h^{\lambda \nu} ) \xi^\mu ] + \frac{ 1 }{ 4 } ( \Delta_\nu h^{\lambda \nu} ) \Delta_\mu \xi^\mu + \frac{ 1 }{ 8 } ( \Delta_\nu h^{\lambda \nu} ) (\Delta_\mu \ln h) \xi^\mu .
\end{equation}

\noindent Then
\begin{equation}                                                           
\Phi [ g ] = \det [ ( - h )^{ - 1 / 8} \tilde{\cal O} ] = \left [ \prod_{\rm sites} ( - h )^{ - 1 / 2} \right ] \det \tilde{\cal O} .
\end{equation}

\noindent Here the first factor $\prod_{\rm sites} ( - h )^{ - 1 / 2}$ changes the measure (\ref{N0-v-alpha-0alpha}) by adding -1 to $\eta$ there. As for $\det \tilde{\cal O}$, it can be represented in the standard way in order to formulate the perturbation theory with the help of auxiliary anticommuting classical fields (Faddeev-Popov ghost fields).
\begin{equation}                                                           
\det \tilde{\cal O} = \int \exp \left ( i \sum_{\rm sites} \otheta_\lambda \tilde{\cal O}^\lambda{}_\mu \vartheta^\mu \right ) \prod_{\rm sites , \nu } \d \otheta_\nu \d \vartheta^\nu ,
\end{equation}

\noindent where $\vartheta^\lambda$, $\otheta_\lambda$ satisfy (here the arguments $x$, $y$ denote sites)
\begin{eqnarray}                                                           
& & \vartheta^\lambda ( x ) \vartheta^\mu ( y ) + \vartheta^\mu ( y ) \vartheta^\lambda ( x ) = 0 , ~~~ \vartheta^\lambda ( x ) \otheta_\mu ( y ) + \otheta_\mu ( y ) \vartheta^\lambda ( x ) = 0 , \nonumber \\ & & \otheta_\lambda ( x ) \otheta_\mu ( y ) + \otheta_\mu ( y ) \otheta_\lambda ( x ) = 0 .
\end{eqnarray}

In overall, the functional integral takes the form
\begin{eqnarray}\label{funct-int-full}                                     
& & \int \exp \{ i ( \tilS_{\rm g} [ g ] + \cF [ g ] + \sum_{\rm sites} \otheta_\lambda \tilde{\cal O}^\lambda{}_\mu \vartheta^\mu ) \} ( \cdot ) \nonumber \\ & & \cdot \prod_{\rm sites} \left ( \prod_\alpha \N_0 (2 \rv_\alpha ) \N_0 (2 \rv_{0 \alpha} ) \right ) ( - g )^\frac{ \eta - 8 }{ 2 } \d^{10} g_{\lambda \mu} \prod_\nu \d \otheta_\nu \d \vartheta^\nu .
\end{eqnarray}

\noindent Here, the first term in $\tilde{\cal O}$ (\ref{tilde-O}) is a finite-difference form of the standard ghost action for the de Donder-Fock gauge, the other two terms are due to the assumed $( - h )^{ - 1 / 4 }$ in $\cF [ g ]$.

\section{Reduction of the functional integral to a form ready for building perturbation series}\label{reduction}

\subsection{Preparation for splitting the measure transformation into one-dimensional problems}\label{splitting}

Now, proceeding to the transformation of the functional measure, we introduce the notation
\begin{eqnarray}\label{mt=f(g)}                                            
& & i \rmm_1 = 2 \rv_1 = i \sqrt{ g_{2 2} g_{3 3} - g^2_{2 3}}, ~ 2 ~ \mbox{perm}(123) , ~~~ \rt_\alpha = 2 \rv_{0 \alpha } = \sqrt{ - g_{0 0} g_{\alpha \alpha } +g^2_{0 \alpha }} , \nonumber \\ & & \rt = \sum_\alpha \rt_\alpha
\end{eqnarray}

\noindent and write the factors of interest as
\begin{eqnarray}\label{N0-exact}                                           
& & \hspace{-10mm} \N_0 (i\rmm_\alpha ) = \frac{\gamma^2}{\gamma^2 + 1} \left | \frac{\rmm^2_\alpha}{\rmm^2_\alpha + 4 \left ( 1 + \frac{i}{\gamma} \right )^{-2} } \right |^2 \left | \frac{2}{1 - \exp \left [ - \pi \left ( 1 + \frac{i}{\gamma} \right ) \rmm_\alpha \right ] } \right |^2 \frac{ e^{ - \pi \rmm_\alpha }}{ \rmm^2_\alpha } , \nonumber \\ & & \hspace{-10mm} \N_0 (\rt_\alpha ) = \frac{\gamma^2}{\gamma^2 + 1} \left | \frac{\rt^2_\alpha}{\rt^2_\alpha + 4 \left ( \frac{1}{\gamma} - i \right )^{-2} } \right |^2 \left | \frac{2}{1 - \exp \left [ - \pi \left ( \frac{1}{\gamma} - i \right ) \rt_\alpha \right ] } \right |^2 \frac{ e^{ - \pi \rt_\alpha / \gamma } }{ \rt^2_\alpha } .
\end{eqnarray}

\noindent Here we keep (for $\rmm_\alpha$,  $\rt_\alpha$ large compared to unity) the leading factors and have up to a constant factor:
\begin{eqnarray}\label{prod-N-0}                                           
\prod_\alpha \N_0 (i\rmm_\alpha ) \N_0 (\rt_\alpha ) & \propto & \prod_\alpha \frac{1}{ \rmm^2_\alpha } \frac{1}{ \rt^2_\alpha } \exp \left ( - \pi \rmm_\alpha - \frac{ \pi }{ \gamma } \rt_\alpha \right ) \nonumber \\ & & = \exp \left ( - \frac{ \pi }{ \gamma } \rt \right ) \prod_\alpha \frac{1}{ \rmm^2_\alpha } \frac{1}{ \rt^2_\alpha } \exp \left ( - \pi \rmm_\alpha \right ) .
\end{eqnarray}

\noindent We also denote the set of the following six metric functions in accordance with
\begin{equation}                                                           
g_{\hlambda \no \hmu} : g_{\hlambda \hmu} = |g_{\lambda \lambda}|^{- 1 / 2} g_{\lambda \mu}|g_{\mu \mu}|^{- 1 / 2} , \lambda \no \mu .
\end{equation}

The idea is to rewrite the measure in terms of $\{ \rmm_\alpha \}$, $\rt$, $\{ g_{\hlambda \no \hmu} \}$ as new variables. Then we pass to a new set $\trmm_\alpha = \trmm_\alpha (\{ \rmm_\beta \}, \rt, \{ g_{\hlambda \no \hmu} \})$, $\trt = \trt (\{ \rmm_\beta \}, \rt, \{ g_{\hlambda \no \hmu} \})$, $\tg_{\hlambda \no \hmu}$ $=$ $g_{\hlambda \no \hmu}$ such that it corresponds to the Lebesgue measure $\d^{10} \tg_{\lambda \mu}$, where $\{ \trmm_\alpha \}$, $\trt$, $\{ \tg_{\hlambda \no \hmu} \}$ are related to $\tg_{\lambda \mu}$ just as $\{ \rmm_\alpha \}$, $\rt$, $\{ g_{\hlambda \no \hmu} \}$ are related to $g_{\lambda \mu}$.

We have for $g = \det \| g_{\lambda \mu} \|$:
\begin{equation}                                                           
g = g_{0 0} g_{1 1} g_{2 2} g_{3 3} \left [ 1 + O ( g^2_{\hlambda \no \hmu} ) \right ] .
\end{equation}

\noindent Here $O ( g^2_{\hlambda \no \hmu} )$ means a value of the order of the terms bilinear in $g_{\hlambda \no \hmu}$. At the same time,
\begin{equation}                                                           
- \prod_\alpha \rmm^2_\alpha \rt^2_\alpha = ( g_{0 0} g_{1 1} g_{2 2} g_{3 3} )^3 \left [ 1 + O ( g^2_{\hlambda \no \hmu} ) \right ] .
\end{equation}

\noindent Therefore,
\begin{equation}                                                           
( - g )^\frac{\eta - 8}{2} = \left [ 1 + O ( g^2_{\hlambda \no \hmu} ) \right ]^\frac{\eta - 8}{6} \left ( \prod_\alpha \rmm_\alpha \rt_\alpha \right )^\frac{\eta - 8}{3} .
\end{equation}

If $\{ \rmm_\alpha \}$, $\{ g_{\hlambda \no \hmu} \}$ are known, then the ratios $\rt_\alpha / \rt$ are known. In particular, if the optimal starting point for the perturbative expansion is reached at $\rmm_\alpha = \rmm^{(0)} \forall \alpha$ (and this is indeed so, as we shall see below), then $\rt_\alpha^{(0)} \forall \alpha$ are the same. Then the geometric mean $(\rt_1 \rt_2 \rt_3 )^{1 / 3}$ coincides with the arithmetic mean $\rt / 3$ of $\rt_1$, $\rt_2$, $\rt_3$ at this point, and their relative difference when shifting from this point in the leading order is bilinear over variations of $\rt_\alpha$ from this point,
\begin{eqnarray}\label{t1t2t3/t^3}                                         
& & \frac{27 \rt_1 \rt_2 \rt_3}{(\rt_1 + \rt_2 + \rt_3)^3} = 1 + \frac{1}{3} \left (\Delta \rt_{\hone} \Delta \rt_{\htwo} + \dots \right ) = 1 + O \left (\Delta \rt^2_{\halpha} \right ) , \nonumber \\ & & \Delta \rt_{\halpha} = \frac{\Delta \rt_\alpha}{\rt_\alpha^{(0)}} = \frac{\rt_\alpha - \rt_\alpha^{(0)}}{\rt_\alpha^{(0)}} .
\end{eqnarray}

\noindent In turn, $\Delta \rt_{\halpha}$ in the leading order linearly depends on $\Delta \rmm_\alpha$ (and on $\Delta g_{00}$, but this is inessential, since in the ratio (\ref{t1t2t3/t^3}), the dependence on $g_{00}$ is cancelled),
\begin{equation}                                                           
\Delta \rt_{\hone} = \frac{1}{2} \left ( \Delta \rmm_{\htwo} + \Delta \rmm_{\hthr} - \Delta \rmm_{\hone} \right ), ~ 2 ~ \mbox{perm}(123) , \Delta \rmm_{\halpha} = \frac{\Delta \rmm_\alpha}{\rmm_\alpha^{(0)}} = \frac{\rmm_\alpha - \rmm_\alpha^{(0)}}{\rmm_\alpha^{(0)}}
\end{equation}

\noindent (we get $\rmm^{(0)}_\alpha = \rmm^{(0)} \forall \alpha$). Therefore,
\begin{equation}\label{t1t2t3/t^3=1+dmdm}                                  
\frac{27 \rt_1 \rt_2 \rt_3}{(\rt_1 + \rt_2 + \rt_3)^3} = 1 + \frac{1}{3} \left (\Delta \rmm_{\hone} \Delta \rmm_{\htwo} + \dots \right ) = 1 + O \left (\Delta \rmm^2_{\halpha} \right ) .
\end{equation}

\noindent Thus,
\begin{equation}\label{g^eta=f(tm)^eta}                                    
( - g )^\frac{\eta - 8}{2} \propto f_g ( g_{\lambda \mu} ) \left (\rt^3 \prod_\alpha \rmm_\alpha \right )^\frac{\eta - 8}{3} , ~~~ f_g ( g_{\lambda \mu} ) = \left [ 1 + O ( g^2_{\hlambda \no \hmu}, \Delta \rmm^2_{\halpha} ) \right ]^\frac{\eta - 8}{6} .
\end{equation}

\noindent Here, in addition to the order of magnitude of the metric variation $O ( g^2_{\hlambda \no \hmu}, \Delta \rmm^2_{\halpha} )$, we also track the dependence on $\eta$ as a potentially large parameter. The scale of the graviton perturbations of the discrete $g_{\lambda \mu}$ (once the coordinate lattice spacing is fixed, $\Delta x^\lambda = 1$) is also related to $\eta$. From the viewpoint of dependence on the scale of the metric, the general form of a term in the action (\ref{Sg[g]}) is similar to
\begin{equation}\label{general-term}                                       
g^{**} g^{**} g^{**} (\Delta_* g_{**}) (\Delta_* g_{**}) (-g)^{1 / 2} \sim g^{**} (\Delta_* g_{**}) (\Delta_* g_{**}) .
\end{equation}

\noindent It turns out that $g_{**}$ has a scale $\sim \eta$, therefore $g^{**} \sim \eta^{- 1}$. Then the scale of graviton metric perturbations $\Delta_* g_{**} \sim \sqrt{\eta }$, and thus $\Delta_* g_{\hast \hast} \sim \eta^{- 1 / 2}$. Then in (\ref{g^eta=f(tm)^eta}) we have
\begin{equation}\label{f=(1+1/eta)^eta}                                    
f_g ( g_{\lambda \mu} ) = \left [ 1 + O ( g^2_{\hlambda \no \hmu}, \Delta \rmm^2_{\halpha} ) \right ]^\frac{\eta - 8}{6} = \left [ 1 + O ( \eta^{- 1} ) \right ]^\frac{\eta - 8}{6} ,
\end{equation}

\noindent whose (binomial) expansion in metric perturbations has no terms growing with $\eta$. In the end, these corrections from metric perturbations will be pumped (by the change of variables considered now) into an expansion of the action in the exponent and further into perturbation theory, as we consider in Subsection \ref{large-parameter}. The above consideration shows that these corrections, moreover, from the factor $( - g )^\frac{\eta - 8}{2}$, which is dangerous in this sense, in the functional measure, will not spoil the perturbative expansion, turning it into an expansion in the large parameter $\eta$.

Returning to the product of $\N_0 (2 \rv_{\lambda \mu} )$ in the measure, we connect the geometric mean of $\rt_1$, $\rt_2$, $\rt_3$ in the same way with their arithmetic mean with relative accuracy $O ( \Delta \rmm^2_{\halpha} )$ and obtain
\begin{eqnarray}\label{prod-N}                                             
& & \prod_\alpha \N_0 (i\rmm_\alpha ) \N_0 (\rt_\alpha ) \propto f_N ( g_{\lambda \mu} ) \frac{1}{\rt^6 } \exp \left ( - \frac{\pi \rt }{\gamma } \right ) \prod_\alpha \frac{1}{ \rmm^2_\alpha } \exp ( - \pi \rmm_\alpha ) , \nonumber \\ & & f_N ( g_{\lambda \mu} ) = 1 + O ( \Delta \rmm^2_{\halpha} ) .
\end{eqnarray}

For the integration element $\d^{10} g_{\lambda \mu}$, we have
\begin{equation}                                                           
\d^{10} g_{\lambda \mu} = 2 \left ( \sqrt{-g_{00}} \right )^4 \left (\d \sqrt{-g_{00}} \right ) \d^3 g_{\hnul \halpha} \d^3 g_{\halpha \no \hbeta} \left ( \prod_\alpha g^{3 / 2}_{\alpha \alpha} \right ) \d^3 g_{\alpha \alpha} ,
\end{equation}

\noindent then we pass from $\{ g_{\alpha \alpha} \}$, $\sqrt{-g_{00}}$ to $\{ \rmm_\alpha \}$, $\rt$, considering $g_{\hlambda \no \hmu}$ as parameters. For $\{ \rmm_\alpha \}$, $\rt$, we have
\begin{equation}                                                           
\rmm_1 = \sqrt{g_{22} g_{33}} \sqrt{1 - g^2_{\htwo \hthr}}, ~ 2 ~ \mbox{perm}(123) , ~ \rt = \sqrt{-g_{00}} \sum_\alpha \sqrt{g_{\alpha \alpha} ( 1 + g^2_{\hnul \halpha} )} .
\end{equation}

\noindent This gives for $\{ g_{\alpha \alpha} \}$, $\sqrt{-g_{00}}$:
\begin{eqnarray}                                                           
g_{11} & = & \frac{\rmm_2 \rmm_3}{\rmm_1} \frac{(1 - g^2_{\htwo \hthr})^{1 / 2}}{(1 - g^2_{\hthr \hone})^{1 / 2} (1 - g^2_{\hone \htwo})^{1 / 2}}, ~ 2 ~ \mbox{perm}(123) , \nonumber \\ \sqrt{-g_{00}} & = & \rt \frac{\prod_{\alpha \no \beta} (1 - g^2_{\halpha \hbeta})^{1 / 4}}{(\rmm_1 \rmm_2 \rmm_3 )^{1 / 2}} \left [ \frac{(1 - g^2_{\htwo \hthr})^{1 / 2} (1 + g^2_{\hnul \hone})^{1 / 2}}{\rmm_1} + ~ 2 ~ \mbox{perm}(123) \right ]^{- 1} \hspace{10mm}
\end{eqnarray}

\noindent and
\begin{eqnarray}                                                           
\d^{10} g_{\lambda \mu} & = & 8 \left \{ \frac{\prod_{\alpha \no \beta} (1 - g^2_{\halpha \hbeta})^{- 3 / 4}}{(\rmm_1 \rmm_2 \rmm_3 )^{5 / 3}} \left [ \frac{(1 - g^2_{\htwo \hthr})^{1 / 2} (1 + g^2_{\hnul \hone})^{1 / 2}}{\rmm_1} \right. \right. \nonumber \\ & & \left. \left. \vphantom{ \frac{(1 - g^2_{\htwo \hthr})^{1 / 2} }{\rmm_1} } + ~ 2 ~ \mbox{perm}(123) \right ]^{- 5} \right \} \d^6 g_{\hlambda \no \hmu} \rt^4 \d \rt \left ( \prod_\alpha \rmm_\alpha \right )^{2 / 3} \d^3 \rmm_\alpha .
\end{eqnarray}

\noindent Here, with a relative accuracy of $O (g^2_{\hlambda \no \hmu})$, the factor in curly brackets is reduced to the (raised to the fifth power) ratio of the harmonic mean of $\rmm_1$, $\rmm_2$, $\rmm_3$ to their geometric mean (which is also the ratio of the geometric mean of $\rmm_1^{-1}$, $\rmm_2^{-1}$, $\rmm_3^{-1}$ to their arithmetic mean) equal to one with an accuracy of $O (\Delta \rmm^2_{\halpha})$,
\begin{equation}\label{d10g}                                               
\d^{10} g_{\lambda \mu} \propto f_0 ( g_{\lambda \mu} ) \d^6 g_{\hlambda \no \hmu} \rt^4 \d \rt \left ( \prod_\alpha \rmm_\alpha^{2 / 3} \right ) \d^3 \rmm_\alpha , ~ f_0 ( g_{\lambda \mu} ) = 1 + O ( g^2_{\hlambda \no \hmu}, \Delta \rmm^2_{\halpha} ).
\end{equation}

The full measure in (\ref{funct-int-full}) (its metric part), consisting of the factors (\ref{g^eta=f(tm)^eta}), (\ref{prod-N}) and (\ref{d10g}), and the equality expressing the fact of its reduction to the Lebesgue measure $\d^{10} \tg_{\lambda \mu}$, similar to (\ref{d10g}), but in the new variables $\{ \trmm_\alpha \}$, $\trt$, takes the form
\begin{eqnarray}\label{f1Dg=f0Dtilde(g)}                                   
& & f_1 ( \{ \rmm_\alpha , g_{\hlambda \no \hmu} \} ) \d^6 g_{\hlambda \no \hmu} \rt^{\eta - 10} \exp (- \pi \rt / \gamma ) \d \rt \left ( \prod_\alpha \rmm_\alpha^\frac{ \eta - 12 }{3} \exp (- \pi \rmm_\alpha) \right ) \d^3 \rmm_\alpha \phantom{\d^3 \trmm_\alpha} \nonumber \\ & & \propto f_0 ( \{ \trmm_\alpha , g_{\hlambda \no \hmu} \} ) \d^6 g_{\hlambda \no \hmu} \trt^4 \d \trt \left ( \prod_\alpha \trmm_\alpha^{2 / 3} \right ) \d^3 \trmm_\alpha, ~~~ f_1 \equiv f_g f_N f_0.
\end{eqnarray}

\noindent $f_1$ and $f_0$ are series over metric perturbations with the leading term 1.

\subsection{Optimal starting point of the perturbative expansion}\label{starting-point}

First consider the optimal starting point of the perturbative expansion. The scale along the coordinate axes is defined by the values $\{ \rmm_\alpha^{(0)} \}$, $\rt^{(0)}$ satisfying the measure maximization condition (\ref{def-l0}). This condition contains $\det \left \| \partial^2 S (\rl_{(0)} ) / (\partial l_i \partial l_k) \right \|$. But $S$ depends bilinearly on the length scale. Indeed, in the given case the role of $S$ is played by the sum $\tilS_{\rm g} + \cF + \sum_{\rm sites} \otheta_\lambda \tilde{\cal O}^\lambda{}_\mu \vartheta^\mu $. Here $\tilS_{\rm g}$ is bilinear in the length scale, and $\cF$ is chosen as such; the ghost action $\sum_{\rm sites} \otheta_\lambda \tilde{\cal O}^\lambda{}_\mu \vartheta^\mu $ is also bilinear in the length scale (and at the initial point $\otheta = 0$, $\vartheta = 0$, its contribution to $\partial^2 S (\rl_{(0)} ) / (\partial l_i \partial l_k)$ disappears altogether). Therefore, if $l_i$ are indeed lengths, then the factor $\det \left \| \partial^2 S (\rl_{(0)} ) / (\partial l_i \partial l_k) \right \|$ does not depend on the length scale. Then we can neglect its influence and look for the maximum of pure $F$. For that, we should temporarily return from $\{ \rmm_\alpha \}$, $\rt$ in the integration element to $\{ \sqrt{g_{\alpha \alpha}} \}$, $\sqrt{-g_{00}}$, which have the meaning of length,
\begin{equation}                                                           
\d^3 \rmm_\alpha \d \rt \propto \left [ 1 + O \left ( g^2_{\hlambda \no \hmu}, \Delta \rmm^2_\halpha \right ) \right ] \prod_\alpha \rmm^{2 / 3}_\alpha \d^3 \sqrt{g_{\alpha \alpha}} \d \sqrt{-g_{00}} .
\end{equation}

\noindent Then the part of the measure depending on $\{ \rmm_\alpha \}$, $\rt$ takes the form
\begin{eqnarray}                                                           
& & \rt^{\eta - 10} \exp \left ( - \frac{\pi \rt }{\gamma } \right ) \d \rt \left ( \prod_\alpha \rmm_\alpha^\frac{\eta - 12}{3 } \exp \left ( - \pi \rmm_\alpha \right ) \right ) \d^3 \rmm_\alpha \propto \left [ 1 + O \left ( g^2_{\hlambda \no \hmu}, \Delta \rmm^2_{\halpha} \right ) \right ]  \nonumber \\ & & \cdot \rt^{\eta - 10} \exp \left ( - \frac{\pi \rt }{\gamma } \right ) \left ( \prod_\alpha \rmm_\alpha^\frac{\eta - 10}{3 } \exp \left ( - \pi \rmm_\alpha \right ) \right ) \d^3 \sqrt{g_{\alpha \alpha}} \d \sqrt{-g_{00}} .
\end{eqnarray}

\noindent It is seen that the maximum of the measure in these coordinates is reached at
\begin{equation}                                                           
\rt = \rt^{(0)} = \gamma \frac{\eta - 10}{ \pi} , ~~~ \rmm_\alpha^{(0)} = \frac{\eta - 10}{3 \pi} \equiv \rmm^{(0)} .
\end{equation}

\noindent Herewith
\begin{equation}                                                           
\rt^{(0)}_\alpha = \gamma \frac{\eta - 10}{3 \pi} = \gamma \rmm_\alpha^{(0)} .
\end{equation}

\noindent The ratio of dynamically given timelike and spacelike areas is equal to $\gamma$.

\subsection{Effective one-dimensional problems}\label{one-dimensional}

Returning to the variables $\{ \rmm_\alpha \}$, $\rt$, we have, in the leading approximation over metric perturbations, the formulas for changing the variables $\{ \rmm_\alpha \} \to \{ \trmm_\alpha \}$, $\rt \to \trt$:
\begin{equation}\label{tmToTilde(tm)}                                      
\rt^{\eta - 10} \exp (- \pi \rt / \gamma) \d \rt \propto \trt^4 \d \trt , ~~~ \rmm_\alpha^\frac{\eta - 12}{3} \exp (- \pi \rmm_\alpha) \d \rmm_\alpha \propto \trmm_\alpha^{2 / 3} \d \trmm_\alpha.
\end{equation}

Then we have to consider the master equation
\begin{equation}\label{int^y=x}                                            
\int^y_0 z^{\tk} \exp (-z ) \d z = x
\end{equation}

\noindent and expand $y = y(x)$ in the vicinity of $y = y_0 = k = \tk + \Delta \tk$ ($\Delta \tk = O(1)$ at large $k$) and $x = x_0$ over $\Delta x = x - x_0$. (In fact, it is the incomplete gamma function $\gamma ( \tk+1, y)$, and we are interested in the reciprocal to it.) Expanding in Taylor series,
\begin{equation}                                                           
y = k + \frac{\d y(x_0)}{\d x} \Delta x + \frac{1}{2} \frac{\d^2 y(x_0)}{\d x^2} \Delta x^2 + \dots ,
\end{equation}

\noindent we have for the derivatives:
\begin{equation}\label{d^ny/dx^n}                                          
\frac{\d^n y(x_0)}{\d x^n} = \left ( e^y y^{- \tk } \frac{\d}{\d y} \right )^{n - 1} e^y y^{- \tk },
\end{equation}

\noindent where we can expand over $\Delta y = y-y_0 = y-k$,
\begin{eqnarray}                                                           
& & e^y y^{- \tk } = e^{k+\Delta y} (k+\Delta y)^{- \tk } = e^k k^{- \tk } e^{\Delta y} \left ( 1+\frac{\Delta y}{k} \right )^{- \tk } = e^k k^{- \tk } \exp \left[ \Delta y \vphantom{\left( 1+\frac{\Delta y}{k} \right)} \right. \nonumber \\ & & \left. - \tk \ln \left( 1+\frac{\Delta y}{k} \right) \right] = e^k k^{- \tk } \exp \left[ \Delta y - \tk \left( \frac{\Delta y}{k } - \frac{1}{2} \frac{\Delta y^2}{k^2 } + \frac{1}{3} \frac{\Delta y^3}{k^3 } - \dots \right) \right] \nonumber \\ & & = e^k k^{- \tk } \exp \left[ \frac{\Delta \tk}{k}  \Delta y + \frac{1}{2} \frac{\tk}{k} \left( \frac{\Delta y}{k^{1 / 2}} \right)^2 - \frac{1}{3} \frac{\tk}{k} \left( \frac{\Delta y}{k^{2 / 3}} \right)^3 + \dots \right] = e^k k^{- \tk } \left\{ 1 \vphantom{\left[ \frac{\tk}{k} \left( \frac{\Delta y}{k^{2 / 3}} \right)^3 + \dots \right]^2} \right. \nonumber \\ & & \left. + \frac{\Delta \tk}{k} \Delta y + \frac{1}{2} \frac{\tk}{k} \left( \frac{\Delta y}{k^{1 / 2}} \right)^2 - \frac{1}{3} \frac{\tk}{k} \left( \frac{\Delta y}{k^{2 / 3}} \right)^3 + \dots + \frac{1}{2!} \left[ \frac{\Delta \tk}{k}  \Delta y \right. \right. \nonumber \\ & & \left. \left. + \frac{1}{2} \frac{\tk}{k} \left( \frac{\Delta y}{k^{1 / 2}} \right)^2 - \frac{1}{3} \frac{\tk}{k} \left( \frac{\Delta y}{k^{2 / 3}} \right)^3 + \dots \right]^2 + \dots \vphantom{\left[ \left( \frac{\Delta y}{k^{1 / 2}} \right)^2 \right]^2} \right\}.
\end{eqnarray}

\noindent In the expansion over $\Delta y$, the coefficient of the largest order of magnitude over $\eta$ at a monomial of $\Delta y$ is $\sim (k \tk^{-1/2})^{-1} = O (\eta^{- 1/2})$ per each factor $\Delta y$, and it is attained for monomials of $\Delta y$ of even degree; for monomials of $\Delta y$ of odd degree $\geq 3$, the largest coefficient is the product of $\sim (k^3 \tk^{-1})^{-1} = O (\eta^{- 2})$ for $\Delta y^3$ and $\sim (k \tk^{-1/2})^{-1} = O (\eta^{- 1/2})$ per each $\Delta y$ for the rest even number of $\Delta y$'s. Thus, differentiation in (\ref{d^ny/dx^n}) leads to such $\d^n  y / \d x^n$ at the point $\Delta y = 0$, which gives the value
\begin{equation}\label{nth-term}                                           
\propto k^{-[n / 2]} \left( e^k k^{-\tk} \Delta x \right)^n \mbox{ at } n \geq 1 \& n \neq 2, ~~~ \propto \frac{\Delta \tk}{k} \left( e^k k^{-\tk} \Delta x \right)^2 \mbox{ at }  n=2
\end{equation}

\noindent for the parameter dependence of the nth term of the Taylor series.

Equality (\ref{int^y=x}) can be replaced by proportionality, and the proportionality constant can be determined so that $x_0$ (chosen arbitrarily) corresponds to $y = y_0 = k$. In addition, we introduce the coefficient $\lambda$ in the exponent $\exp ( - \lambda z )$ (the initial point is $y_0 = k / \lambda$).
\begin{equation}\label{int^y=Cx}                                           
\int^y_0 z^\tk \exp (- \lambda z) \d z = C x, ~~~ C = x_0^{- 1} \lambda^{- \tk - 1} \int^k_0 z^\tk \exp (- \lambda z) \d z.
\end{equation}

\noindent Then the nth term (\ref{nth-term}) in the Taylor series expansion of $y$ in $\Delta x$ becomes
\begin{eqnarray}\label{nth-term1}                                          
& & \propto \lambda^{n - 1} k^{- [n / 2]} \left [ e^k \left ( \frac{k}{\lambda} \right)^{-\tk} \int^k_0 z^\tk \exp (- z) \d z ~ \lambda^{- \tk - 1} \frac{\Delta x}{x_0} \right ]^n \nonumber \\ & & = \lambda^{-1} k^{- [n / 2]} \left( e^k k^{- \tk} \int^k_0 z^\tk \exp (- z) \d z \frac{\Delta x}{x_0} \right)^n = \lambda^{-1} k^{- [n / 2]} \left( \sqrt{\frac{ \pi k}{2 \tk}} \sqrt{k} \frac{\Delta x}{x_0} \right)^n \hspace{5mm} \nonumber \\ & & = \lambda^{-1} k^{n/2 - [n / 2]} \left( \sqrt{\frac{ \pi k}{2 \tk}} \frac{\Delta x}{x_0} \right)^n \mbox{ at } n \geq 1 \& n \neq 2, \nonumber \\ & & = \frac{\pi k \Delta \tk}{4 \lambda \tk} \left( \frac{\Delta x}{x_0} \right)^2 \mbox{ at }  n=2, ~~~ = \frac{k}{\lambda} \mbox{ at }  n=0.
\end{eqnarray}

\subsection{Absence of a large parameter in the perturbative expansion}\label{large-parameter}

For the transitions $\rt \to \trt$ and $\rmm_\alpha \to \trmm_\alpha$, we have
\begin{eqnarray}                                                           
& & \tk = k = \eta - 10, ~~~ \lambda = \frac{\pi }{\gamma }, ~~~ x = \frac{\trt^5}{5} ~~~ \mbox{ and } \nonumber \\ & & \tk = \frac{\eta - 12}{3}, ~~~ k = \frac{\eta - 10}{3}, ~~~ \lambda = \pi, ~~~ x = \frac{3}{5} \trmm_\alpha^{5/3},
\end{eqnarray}

\noindent respectively. A more detailed explicit calculation gives for these transitions
\begin{eqnarray}\label{t,m=f(tilde(t,m))}                                  
\rt & = & \gamma \frac{\eta - 10}{\pi} + \gamma \sqrt{\frac{\eta - 10}{2 \pi}} \frac{\Delta (\trt^5 )}{\trt^{(0) 5}} + \gamma \frac{\pi}{12 } \sqrt{\frac{\eta - 10}{2 \pi}} \left[ \frac{\Delta (\trt^5 )}{\trt^{(0) 5}} \right]^3 + \dots, \nonumber \\ \rmm_\alpha & = & \frac{\eta - 10}{3 \pi} + \frac{\eta - 10}{\sqrt{6\pi (\eta - 12)}} \frac{\Delta (\trmm_\alpha^{5/3})}{\trmm_\alpha^{(0) 5/3}} + \frac{1}{6} \frac{\eta - 10}{\eta - 12} \left[ \frac{\Delta (\trmm_\alpha^{5/3})}{\trmm_\alpha^{(0) 5/3}} \right]^2 \nonumber \\ & & + \frac{\pi }{36 } \frac{3 \eta - 28}{\eta - 12} \frac{\eta - 10}{\sqrt{6 \pi (\eta - 12)}} \left[ \frac{\Delta (\trmm_\alpha^{5/3})}{\trmm_\alpha^{(0) 5/3}} \right]^3 + \dots.
\end{eqnarray}

\noindent Here $x_0^{-1} \Delta x$, in turn, expands into a binomial expansion,
\begin{eqnarray}\label{Dx/x0}                                              
\frac{\Delta (\trt^5 )}{\trt^{(0) 5}} = 5 \frac{\Delta (\trt )}{\trt^{(0)}} + 10 \left( \frac{\Delta (\trt )}{\trt^{(0)}} \right)^2 + \dots, \nonumber \\ \frac{\Delta (\trmm_\alpha^{5/3})}{\trmm_\alpha^{(0) 5/3}} = \frac{5}{3} \frac{\Delta \trmm_\alpha}{\trmm_\alpha^{(0)}} + \frac{5}{9} \left( \frac{\Delta \trmm_\alpha}{\trmm_\alpha^{(0)}} \right)^2 + \dots.
\end{eqnarray}

\noindent If the resulting expansions for $\rt$, $\{ \rmm_\alpha \}$ are substituted into the action, the scale of graviton perturbations is given by the bilinear form in the action for the linear terms $\Delta \trt / \trt^{(0)}$, $\Delta \trmm_\alpha / \trmm_\alpha^{(0)}$ in $\rt$, $\{ \rmm_\alpha \}$. These terms appear in $\rt$, $\{ \rmm_\alpha \}$ already with coefficients $O(\sqrt{\eta})$, so the dynamics forces them to be $O(1)$ (cf. considering the scale of graviton perturbations based of the general term in the action (\ref{general-term})). Then $\Delta (\trt^5 )/\trt^{(0) 5}$, $\Delta (\trmm_\alpha^{5/3})/\trmm_\alpha^{(0) 5/3}$ (\ref{Dx/x0}) are $O(1)$ as well. Then in the expansions for $\rt$, $\{ \rmm_\alpha \}$ (\ref{t,m=f(tilde(t,m))}), the terms $n \geq 1$ are either $O(\sqrt{\eta})$ or $O(1)$, but not of higher order in $\eta$, which would generate vertices and diagrams that make up an expansion in increasing powers of the large parameter $\eta$.

\subsection{Typical correction to the splitting approximation}\label{correction}

Earlier, we made a similar conclusion on the absence of growing powers of $\eta$ in the multiplicative correction $f_g ( g_{\lambda \mu} )$ (due to metric perturbations) to the expression used for $( - g )^\frac{\eta - 8}{2}$, (\ref{g^eta=f(tm)^eta}). (This is all the more true for other correction functions $f_N$ and $f_0$ (\ref{f1Dg=f0Dtilde(g)}), in which there is no explicit dependence on $\eta$.) Having got the basic transformation $\rt \to \trt$, $\rmm_\alpha \to \trmm_\alpha$, we can take into account such corrections with the help of small additional transformations. For example, consider the relative correction $\sim \Delta \rmm_{\hone} \Delta \rmm_{\htwo}$ (\ref{t1t2t3/t^3=1+dmdm}), which appears in the correction function $f_g$ as
\begin{equation}                                                           
f_g = \left( 1 + \frac{1}{3} \Delta \rmm_{\hone} \Delta \rmm_{\htwo} + \dots \right)^\frac{\eta - 8}{3} = 1 + \frac{\eta - 8}{9} \Delta \rmm_{\hone} \Delta \rmm_{\htwo} + \dots .
\end{equation}

\noindent Taking into account this correction modifies the formulas for changing the variables $\{ \rmm_\alpha \} \to \{ \trmm_\alpha \}$, $\rt \to \trt$ (\ref{tmToTilde(tm)})  in terms of the variables $\rmm_1$, $\rmm_2$:
\begin{eqnarray}\label{(1+dmdm)Dm=Dtilde(m)}                               
\left( 1 + \frac{\eta - 8}{9} \frac{\Delta \rmm_1}{\rmm^{(0)}} \frac{\Delta \rmm_2}{\rmm^{(0)}}  + \dots \right) \rmm_1^\frac{\eta - 12}{3} e^{- \pi \rmm_1} \d \rmm_1 \rmm_2^\frac{\eta - 12}{3} e^{- \pi \rmm_2} \d \rmm_2 \nonumber \\ \propto \d ( \trmm_1^{5/3} ) \d ( \trmm_2^{5/3} ).
\end{eqnarray}

\noindent Here we substitute the above solution for $\rt$, $\{ \rmm_\alpha \}$ in terms of $\trt$, $\{ \trmm_\alpha \}$ (\ref{t,m=f(tilde(t,m))}), where we now relabel $\trt$, $\{ \trmm_\alpha \}$ as $\ttrt$, $\{ \ttrmm_\alpha \}$ (intermediate variables),
\begin{eqnarray}\label{m=m(0)+m(1)}                                        
\rmm_\alpha = \frac{\eta - 10}{3 \pi} + \frac{\eta - 10}{\sqrt{6\pi (\eta - 12)}} \frac{\Delta (\ttrmm_\alpha^{5/3})}{\ttrmm_\alpha^{(0) 5/3}} + \dots, \nonumber \\ \frac{\Delta \rmm_\alpha}{\rmm^{(0)}} = \sqrt{\frac{3 \pi}{2 (\eta - 12)}} \frac{\Delta (\ttrmm_\alpha^{5/3})}{\ttrmm_\alpha^{(0) 5/3}} + \dots,
\end{eqnarray}

\noindent and find a condition on the additional transformation $\{ \ttrmm_\alpha \} \to \{ \trmm_\alpha \}$,
\begin{equation}\label{(1+DmDm)dmdm=dmdm}                                  
\hspace{-3mm}\left( 1 + \frac{\pi}{6} \frac{\eta - 8}{\eta - 12} \frac{\Delta (\ttrmm_1^{5/3})}{\ttrmm_1^{(0) 5/3}}  \frac{\Delta (\ttrmm_2^{5/3})}{\ttrmm_2^{(0) 5/3}} + \dots \right) \frac{\d (\ttrmm_1^{5/3})}{\ttrmm_1^{(0) 5/3}}  \frac{\d (\ttrmm_2^{5/3})}{\ttrmm_2^{(0) 5/3}} = \frac{\d (\trmm_1^{5/3})}{\trmm_1^{(0) 5/3}}  \frac{\d (\trmm_2^{5/3})}{\trmm_2^{(0) 5/3}},
\end{equation}

\noindent where the proportionality constant is chosen so that in the absence of corrections, $\ttrmm_\alpha = \trmm_\alpha$ can be chosen. Here the dots (higher corrections) are combined from the dots in (\ref{(1+dmdm)Dm=Dtilde(m)}) and the dots in (\ref{m=m(0)+m(1)}). (Of course, if the correction encountered depends on more variables, the corresponding integration elements should be taken into account on both sides of the equation.) We can choose
\begin{eqnarray}                                                           
& & \frac{\Delta (\ttrmm_1^{5/3})}{\ttrmm_1^{(0) 5/3}} = \frac{\Delta (\trmm_1^{5/3})}{\trmm_1^{(0) 5/3}} - \frac{\pi }{12 } \frac{\eta - 8}{\eta - 12} \left( \frac{\Delta (\trmm_1^{5/3})}{\trmm_1^{(0) 5/3}} \right)^2 \frac{\Delta (\trmm_2^{5/3})}{\trmm_2^{(0) 5/3}} + \dots , \nonumber \\ & & \frac{\Delta (\ttrmm_2^{5/3})}{\ttrmm_2^{(0) 5/3}} = \frac{\Delta (\trmm_2^{5/3})}{\trmm_2^{(0) 5/3}} + \dots,
\end{eqnarray}

\noindent or, more symmetrically,
\begin{eqnarray}\label{Dm=Dm-Dm2DmSymm}                                    
& & \frac{\Delta (\ttrmm_1^{5/3})}{\ttrmm_1^{(0) 5/3}} = \frac{\Delta (\trmm_1^{5/3})}{\trmm_1^{(0) 5/3}} - \frac{\pi }{24 } \frac{\eta - 8}{\eta - 12} \left( \frac{\Delta (\trmm_1^{5/3})}{\trmm_1^{(0) 5/3}} \right)^2 \frac{\Delta (\trmm_2^{5/3})}{\trmm_2^{(0) 5/3}} + \dots , \nonumber \\ & & \frac{\Delta (\ttrmm_2^{5/3})}{\ttrmm_2^{(0) 5/3}} = \frac{\Delta (\trmm_2^{5/3})}{\trmm_2^{(0) 5/3}} - \frac{\pi }{24 } \frac{\eta - 8}{\eta - 12} \frac{\Delta (\trmm_1^{5/3})}{\trmm_1^{(0) 5/3}} \left( \frac{\Delta (\trmm_2^{5/3})}{\trmm_2^{(0) 5/3}} \right)^2 + \dots
\end{eqnarray}

\noindent (in this case, if only the shown terms are used, equality (\ref{(1+DmDm)dmdm=dmdm}) is violated by a relative amount of the order of $O \left (\Delta \ttrmm^4_{\halpha} \right )$, but higher relative corrections $O \left (\Delta \ttrmm^4_{\halpha} \right )$ in (\ref{Dm=Dm-Dm2DmSymm}) should compensate for this). As above, since the dynamics forces $\Delta (\trmm_\alpha^{5/3}) /\trmm_\alpha^{(0) 5/3}$ to be $O(1)$, the correction found is $O(1)$ as well. This illustrates converting corrections to the measure to corrections to the parametrization of $g_{\lambda \mu}$ in terms of $\tg_{\lambda \mu}$.

Together with the above consideration of the expansion of $( - g )^\frac{\eta - 8}{2}$ over metric perturbations (\ref{g^eta=f(tm)^eta}), (\ref{f=(1+1/eta)^eta}) and the consideration of the expansion of $\rt$, $\{ \rmm_\alpha \}$ in Subsection \ref{large-parameter}, this allows us to complete checking the absence of a large parameter in the perturbative expansion when substituting $g_{\lambda \mu} = g_{\lambda \mu} ( \tg_{\lambda \mu} )$ into the action.

\subsection{Specificity of the maximum point of the functional measure in the aspect of perturbative expansion}\label{maximum}

At the same time, we note that there would be no such absence of a large parameter in the perturbative expansion if the initial point were far enough from the maximum point of the measure. To see this, let us pay attention to the coefficient of the second-order term ($n = 2$ in (\ref{nth-term1})) in the Taylor series expansion of $y$ in $x$ around $y = k$, which solves the master equation (\ref{int^y=Cx}) (at $\lambda = 1$), which reduces the measure $e^{- y} y^\tk \d y$ to the Lebesgue one, $\d x$. If $\Delta \tk (= k - \tk) = O ( \tk^p )$, where $p > 1/2$, then this term generates vertices and diagrams that make up an expansion in increasing powers of the large parameter $\tk$ (that is, $\eta$).

\begin{figure}[h]
	\centering
	\includegraphics[scale=1]{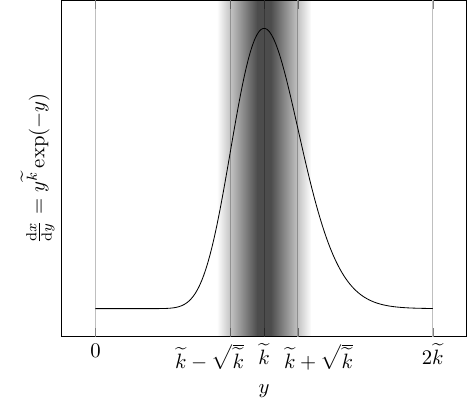}
	\caption{An interval of the initial points $y$ that do not lead to a large parameter in the perturbative expansion. $y$ has the meaning of an area variable, $\d x ( y )$ is a measure. The fuzziness of the boundaries corresponds to the order-of-magnitude accuracy of estimating their displacement from the maximum point of the measure given in the text.}
	\label{f1}
\end{figure}

That is, the initial points $y = k$ that do not lead to a large parameter in the perturbative expansion are contained in the interval $[\tk - O ( \tk^{1 / 2}), \tk + O ( \tk^{1 / 2})]$ (Fig.~\ref{f1}).
\indent It is tempting to attribute the boundary points of the interval to the inflection points of the function $e^{- y} y^\tk$, i.e., $\tk \pm \tk^{1 / 2}$. However, these pairs of points are defined by somehow related but not identical conditions: the inflection points are defined by the vanishing of the second derivative, and $k$ outside the interval is defined by a sufficient value of the logarithmic derivative such that $(e^{- y} y^\tk )^{- 1} (e^{- y} y^\tk )^\prime = - k^{- 1} \Delta \tk$ is of order $O(\tk^{p -1})$ in absolute value at $p>1/2$.

A particular limiting case is the case when there is no exponential suppression of the scale of metric variables along some axis. For example, let us omit the factors $\N_0 (2 \rv_{0 \alpha} )$ in the expression for measure (\ref{N0-v-alpha-0alpha}), which suppress the contribution of large timelike areas. Then we have the task of transforming the measure $y^\tk \d y$: $y^\tk \d y = C \d x$, where $\tk = \eta - 4$, $y = \sqrt{- g_{0 0}}$. Expanding around some point $y = k$ and choosing $C = k^{\tk + 1 / 2}$, we obtain
\begin{equation}\label{y=x^(1/(k+1))}                                      
y = k \left[ 1 + \frac{\tk + 1}{\sqrt{k}} \Delta x \right]^\frac{1}{\tk + 1} = k + \sqrt{k} \Delta x - \frac{1}{2} \tk \Delta x^2 + \frac{1}{2} \tk \frac{2 \tk + 1}{\sqrt{k}} \Delta x^3 + \dots .
\end{equation}

\noindent The linear in $\Delta x$ term takes part in forming the bilinear form of the action, and the dynamics forces typical graviton perturbations $\Delta x$ to be of order $O(1)$ (cf. considering the scale of graviton perturbations based of the general term in the action (\ref{general-term}); in the absence of a dynamically specified scale $k$ of $y$, we take $k = O ( \tk )$ to match the main text, where $k = \tk + O ( 1 )$). Then the higher order terms have orders of increasing powers of $\tk$. Each of these terms will generate vertices and diagrams that make up an expansion in increasing powers of the large parameter $\eta$.

In this case, we see that the matter is not only in the second derivative of the function $y (x )$ inverse to the measure (or in the first derivative of $\d x / \d y$, i.e., of the measure itself), but also in higher ones. So it is fortunate that, as shown above, the dependence on $\eta$ of {\it each} derivative of the function $y(x)$ in the vicinity of the maximum point of the measure is limited to the order $O(\sqrt{\eta })$.

\subsection{Parametrization of the metric up to bilinear terms}\label{parametrization}

Now we write out parametrization of the metric $g_{\lambda \mu}$ in terms of $\tg_{\lambda \mu}$ up to bilinear terms.

Metric-bilinear relative corrections to the measure lead to cubic corrections to the parametrization of $g_{\lambda \mu}$ in terms of $\tg_{\lambda \mu}$, which in turn leads to 4-graviton vertices. In diagrams for quantities such as Newton's potential, 4-graviton vertices appear starting with two-loop diagrams.

In contrast, bilinear corrections to the parametrization of the metric lead to 3-graviton vertices that appear in the diagrams for the Newtonian potential already at the level of one loop. Moreover, the set of such corrections is combinatorially much simpler: these corrections do not arise from corrections to the measure, but appear only in the dependence of $\rt$, $\{ \rmm_\alpha \}$ on $\trt$, $\{ \trmm_\alpha \}$ (\ref{t,m=f(tilde(t,m))},\ref{Dx/x0}).

The second order over $\Delta \tg_{\lambda \mu}$ arises, by means of (\ref{Dx/x0}), from the second terms in $\rt$, $\{ \rmm_\alpha \}$ and the third term $[\Delta (\trmm_\alpha^{5/3})/\trmm_\alpha^{(0) 5/3}]^2$ in $\{ \rmm_\alpha \}$ in (\ref{t,m=f(tilde(t,m))}). We express $\trt$, $\{ \trmm_\alpha \}$ in terms of $\tg_{\lambda \mu}$ according to
\begin{equation}                                                           
\trmm_1 = \sqrt{ \tg_{2 2} \tg_{3 3}(1 - \tg^2_{\htwo \hthr})}, ~ 2 ~ \mbox{perm}(123) , ~~~ \trt = \sum_\alpha \sqrt{ - \tg_{0 0} \tg_{\alpha \alpha }(1 + \tg^2_{\hnul \halpha })} .
\end{equation}

\noindent We choose $\trmm^{(0)}_\alpha = 1 \forall \alpha$, $\trt^{(0)} / 3 = 1$, and thus $\tg^{(0)}_{\lambda \mu} = \eta_{\lambda \mu}$, $\Delta \tg_{\lambda \mu} = \tg_{\lambda \mu} - \eta_{\lambda \mu}$, so
\begin{eqnarray}\label{t,m=f(tilde(g))}                                    
\rt & = & \gamma \frac{\eta - 10}{\pi} + 5 \gamma \sqrt{\frac{\eta - 10}{2 \pi}} \left[ - \frac{1}{2} \Delta \tg_{0 0} + \frac{1 }{6 } \sum_\alpha \Delta \tg_{\alpha \alpha} + \frac{1 }{2 } \left( \Delta \tg_{0 0} - \frac{1}{3} \sum_\alpha \Delta \tg_{\alpha \alpha} \right)^2 \right. \nonumber \\ & & \left. - \frac{1}{24 } \sum_\alpha \left( \Delta \tg_{0 0} + \Delta \tg_{\alpha \alpha} \right)^2 + \frac{1}{6} \sum_\alpha \tg^2_{\hnul \halpha} \right] + \dots , \nonumber \\ \rmm_1 & = & \frac{\eta - 10}{3 \pi} + \frac{5 }{6 } \frac{\eta - 10}{\sqrt{6\pi (\eta - 12)}} \left[ \Delta \tg_{2 2} + \Delta \tg_{3 3} - \frac{1}{12 } \Delta \tg^2_{2 2} - \frac{1}{12 } \Delta \tg^2_{3 3} \right. \nonumber \\ & & \left. + \frac{5}{6} \Delta \tg_{2 2} \Delta \tg_{3 3} - \tg^2_{\htwo \hthr} \right] + \frac{25}{216} \frac{\eta - 10}{\eta - 12} (\Delta \tg_{2 2} + \Delta \tg_{3 3})^2 + \dots, ~ 2 ~ \mbox{perm}(123).
\end{eqnarray}

Then we can write (\ref{t,m=f(tilde(g))}) as equalities for $g_{\lambda \lambda}$ instead of $\rt$, $\{ \rmm_\alpha \}$. Now we also use the relation $g_{\hlambda \no \hmu} = \tg_{\hlambda \no \hmu}$ to write such equalities for $g_{\lambda \no \mu}$ as well,
\begin{eqnarray}\label{g(lm)=1+Dtilde(g)}                                  
& & \frac{3 \pi \gamma^{-2}}{ \eta - 10} g_{0 0} = - 1 + 5 \sqrt{ \frac{ \pi }{2 (\eta - 10)}} \left\{ \Delta \tg_{0 0} - \frac{1}{3} \left(1 - \sqrt{\frac{\eta - 10}{3 (\eta - 12)}}  \right) \sum_\alpha \Delta \tg_{\alpha \alpha} \right. \nonumber \\ & & \left. - \frac{3}{4} \left[ \Delta \tg_{0 0} - \frac{1}{3} \left(1 - \sqrt{\frac{\eta - 10}{3 (\eta - 12)}}  \right) \sum_\alpha \Delta \tg_{\alpha \alpha} \right]^2 + \left(\frac{1 }{3 } + \frac{1 }{2 } \sqrt{\frac{\eta - 10}{3 (\eta - 12)}} \right) \right. \nonumber \\ & & \left. \cdot \left[ \Delta \tg_{0 0} - \frac{1}{3} \left(1 - \sqrt{\frac{\eta - 10}{3 (\eta - 12)}} \right) \sum_\alpha \Delta \tg_{\alpha \alpha} \right] \sum_\alpha \Delta \tg_{\alpha \alpha} + \frac{1}{6} \left( \frac{\eta - 13}{3 (\eta - 12)} \right. \right. \nonumber \\ & & \left. \left. - \frac{1}{4 } \sqrt{\frac{\eta - 10}{3 (\eta - 12)}} \right) \left(\sum_\alpha \Delta \tg_{\alpha \alpha} \right)^2 + \left(\frac{1 }{12 } - \frac{7 }{72 } \sqrt{\frac{\eta - 10}{3 (\eta - 12)}} \right) \sum_\alpha \Delta \tg^2_{\alpha \alpha} \right\} \nonumber \\ & & + \frac{1 }{3 } \left( 1 - 5 \sqrt{ \frac{ \pi }{2 (\eta - 10)}} \right) \sum_\alpha \tg^2_{\hnul \halpha} + \frac{1 }{6 } \left( 1 - 5 \sqrt{ \frac{ \pi }{6 (\eta - 12)}} \right) \sum_{\alpha \no \beta} \tg^2_{\halpha \hbeta} \nonumber \\ & & + \frac{25 }{216} \frac{\pi }{\eta - 12} \left[ 2 \left( \sum_\alpha \Delta \tg_{\alpha \alpha} \right)^2 + \sum_\alpha \Delta \tg^2_{\alpha \alpha} \right] , \nonumber \\ & & \frac{3 \pi }{ \eta - 10} g_{1 1} = 1 + 5 \sqrt{ \frac{ \pi }{6 (\eta - 12)}} \left[ \Delta \tg_{1 1} - \frac{1}{12} \Delta \tg^2_{1 1} + \frac{5}{12} \left( \Delta \tg_{3 3} \tg_{1 1} + \Delta \tg_{1 1} \tg_{2 2} \right. \right. \nonumber \\ & & \left. \left. - \Delta \tg_{2 2} \tg_{3 3} \right) \vphantom{\frac{5 }{12 }} \right] + \left[ \frac{1}{2} - \frac{5}{2 } \sqrt{ \frac{ \pi }{6 (\eta - 12)}} \right] \left( \tg^2_{\hthr \hone} + \tg^2_{\hone \htwo} - \tg^2_{\htwo \hthr} \right) + \frac{25 }{72 } \frac{\pi }{ \eta - 12 } \nonumber \\ & & \cdot \left[ \left( \Delta \tg_{33} + \Delta \tg_{11} \right)^2 + \left( \Delta \tg_{11} + \Delta \tg_{22} \right)^2 + 2 \left( \Delta \tg_{22} + \Delta \tg_{33} \right)^2 \right] , ~ 2 ~ \mbox{perm}(123), \nonumber \\ & & \frac{3 \pi \gamma^{-1}}{ \eta - 10} g_{0 \alpha} = \tg_{0 \alpha} + \frac{1}{2} \left[ \Delta \tg_{0 0} - \Delta \tg_{\alpha \alpha} - 5 \sqrt{\frac{\pi }{2 (\eta - 10)}} \left( \Delta \tg_{0 0} - \frac{1}{3} \sum_\beta \Delta \tg_{\beta \beta} \right) \right. \nonumber \\ & & \left.+ 5 \sqrt{\frac{\pi }{6 (\eta - 12)}} \left( \Delta \tg_{\alpha \alpha} - \frac{1}{3} \sum_\beta \Delta \tg_{\beta \beta} \right) \right] \tg_{0 \alpha} , \nonumber \\ & & \frac{3 \pi }{ \eta - 10} g_{\alpha \beta} = \tg_{\alpha \beta} - \frac{1}{2} \left( 1 - 5 \sqrt{ \frac{ \pi }{6 (\eta - 12)}} \right) \left( \Delta \tg_{\alpha \alpha} + \Delta \tg_{\beta \beta} \right) \tg_{\alpha \beta}, ~~~ \alpha \neq \beta .
\end{eqnarray}

\noindent In the formulas $g_{\lambda \mu} = g_{\lambda \mu} ( \{ \tg_{\nu \rho} \} )$, the dependence on the parameter $\gamma$ is reduced to a certain power factor at $g_{\lambda \mu}$. This is due to the use of the approximate expression for $\N_0 (\rt_\alpha )$ (\ref{prod-N-0}), where, up to a power factor, the dependence on $\gamma$ can be absorbed by $\rt_\alpha$: $\N_0 ( \gamma \rt_\alpha ) = \N_0 (\rt_\alpha ) |_{\gamma = 1} \gamma^{- 2}$. In larger orders over $\rt^{- 1}$, i.e. over $\eta^{- 1}$, as can be seen from (\ref{N0-exact}), this property is violated.

To formulate in near-continuum notation, we pass from the metric $g_{\lambda \mu}$ with the help of some auxiliary constant tetrad $l^{(0) a}_\lambda$ to a new metric $\rg_{a b}$ for the rescaled coordinate axes,
\begin{eqnarray}\label{g_(lm)=lg_(ab)l}                                    
& & g_{\lambda \mu} = l^{(0) a}_\lambda \rg_{a b} l^{(0) b}_\mu, ~~~ l^{(0) a}_\lambda = \sqrt{\frac{\eta - 10}{3 \pi}} {\rm diag}(\gamma,1,1,1), \nonumber \\ & & \rg_{a b} = \eta_{a b} + w_{a b} + B_{\srg_{a b}}^{w_{c d} w_{e f}} w_{c d} w_{e f} + \dots.
\end{eqnarray}

\noindent Here $w_{a b}$ is the part of $\rg_{a b}$ that is linear in $\Delta \tg_{\lambda \mu}$. Linear transformations of $\tg_{\lambda \mu}$ as parametrization variables are allowed, since they do not change, up to proportionality, the form of the measure $\d^{10} \tg_{\lambda \mu}$. In $\rg_{a b}$, the matrix $w_{a b}$ can be taken as a new matrix variable instead of $\tg_{\lambda \mu}$. The above calculation gives $O (\sqrt{\eta}))$ for the matrix $B$ defining the bilinear term. For simplicity, we present the result of going from $\Delta \tg_{\lambda \mu}$ to $w_{a b}$ in (\ref{g(lm)=1+Dtilde(g)}) in this leading order over $\eta$. More precisely, we keep the difference between $\eta - 10$ and $\eta - 12$ to indicate the different origin of the corresponding terms. The resulting bilinear term is given by

\begin{eqnarray}\label{g_(ab)}                                             
& & \frac{3 \pi \gamma^{-2}}{ \eta - 10} g_{0 0} = - 1 + w_{0 0} + \frac{\sqrt{6}}{5} \sqrt{\frac{\eta - 12}{\pi }} \left\{ - \frac{\sqrt{3}}{4} \sqrt{\frac{\eta - 10}{\eta - 12}} w^2_{0 0} + \frac{\sqrt{3} + 2}{6} w_{0 0} \sum_\alpha w_{\alpha \alpha} \right. \nonumber \\ & & \left. + \sqrt{\frac{\eta - 12}{\eta - 10}} \left[ \frac{4 \sqrt{3} - 3}{72} \left( \sum_\alpha w_{\alpha \alpha} \right)^2 + \frac{6 \sqrt{3} - 7}{72} \sum_\alpha w^2_{\alpha \alpha} \right] \right\} , \nonumber \\ & & \frac{3 \pi }{ \eta - 10} g_{1 1} = 1 + w_{1 1} + \frac{\sqrt{6}}{5} \sqrt{\frac{\eta - 12}{\pi }} \left[ - \frac{1}{12} w^2_{1 1} + \frac{5}{12} \left( w_{3 3} w_{1 1} + w_{1 1} w_{2 2} \right. \right. \nonumber \\ & & \left. \left. - w_{2 2} w_{3 3} \right) \vphantom{\frac{5}{12 }} \right], ~ 2 ~ \mbox{perm}(123), \nonumber \\ & & \frac{3 \pi \gamma^{-1}}{ \eta - 10} g_{0 \alpha} = w_{0 \alpha} + \frac{\sqrt{6}}{10} \sqrt{\frac{\eta - 12}{\pi }} \left[ \sqrt{\frac{\eta - 10}{3 (\eta - 12)}} \left( w_{0 0} - \frac{1}{3} \sum_\beta w_{\beta \beta} \right) - w_{\alpha \alpha} \right. \nonumber \\ & & \left. + \frac{1}{3}  \sum_\beta w_{\beta \beta} \right] w_{0 \alpha} , \nonumber \\ & & \frac{3 \pi }{ \eta - 10} g_{\alpha \beta} = w_{\alpha \beta} - \frac{\sqrt{6}}{10} \sqrt{\frac{\eta - 12}{\pi }} \left( w_{\alpha \alpha} + w_{\beta \beta} \right) w_{\alpha \beta}, ~~~ \alpha \neq \beta .
\end{eqnarray}

\subsection{Propagators}\label{propagators}

Substituting $g_{\lambda \mu}$ in terms of $\rg_{a b}$ (\ref{g_(lm)=lg_(ab)l}) into the action (\ref{Sg[g]}) gives
\begin{eqnarray}\label{Sg[LgL^(-1)]}                                       
& & \tilS_{\rm g} [ g ] = \frac{1}{ 8 } \sum_{\rm sites} ( \tDelta_c \rg_{a b} ) ( \tDelta_f \rg_{d e} ) ( 2 \rg^{a d} \rg^{b f} \rg^{c e} - \rg^{a d} \rg^{b e} \rg^{c f} - 2 \rg^{a f} \rg^{b c} \rg^{d e} \nonumber \\ & & + \rg^{a b} \rg^{d e} \rg^{c f} ) \sqrt{ - \det \| \rg_{a b} \| }, \mbox{ ~ where } \tDelta_a = \sqrt{ \det \| l^{(0) b}_\mu \|} l^{(0) \lambda}_a \Delta_\lambda \equiv \tl^{(0) \lambda}_a \Delta_\lambda
\end{eqnarray}

\noindent ($l^{(0) \lambda}_a$ is the reciprocal to $l^{(0) a}_\lambda$).

The $\alpha$-term (\ref{alpha-term}) takes the form
\begin{eqnarray}\label{alpha-term-g}                                       
& & \cF [ g ] = - \frac{\alpha }{4 } \sum_{\rm sites} [\tDelta_c ( \rg^{a c} \sqrt{- \rg} )] (- \rg)^{- 1 / 4} (\det \| \tl_f^{(0) \sigma} \|)^{- 1 / 2} \tl_a^{(0) \lambda} \teta_{\lambda \mu} \tl_b^{(0) \mu} \nonumber \\ & & \cdot [\tDelta_d ( \rg^{b d} \sqrt{- \rg} )] = - \frac{\alpha }{4 } \sum_{\rm sites} [\tDelta_c ( \rg^{a c} \sqrt{- \rg} )] (- \rg)^{- 1 / 4} \eta_{a b} [\tDelta_d ( \rg^{b d} \sqrt{- \rg} )]
\end{eqnarray}

\noindent ($\rg \equiv \det \| \rg_{a b} \|$), where to get
\begin{equation}                                                           
(\det \| \tl_f^{(0) \sigma} \|)^{- 1 / 2} \tl_a^{(0) \lambda} \teta_{\lambda \mu} \tl_b^{(0) \mu} = \eta_{a b}
\end{equation}

\noindent we should take
\begin{equation}\label{tilde-eta}                                          
\teta_{\lambda \mu} = (\det \| \tl_f^{(0) \sigma} \|)^{1 / 2} \tl_\lambda^{(0) a} \eta_{a b} \tl_\mu^{(0) b} = {\rm diag}(- \gamma^{3 / 2},\gamma^{- 1 / 2},\gamma^{- 1 / 2},\gamma^{- 1 / 2})
\end{equation}

\noindent ($\tl_\lambda^{(0) a}$ is the reciprocal to $\tl_a^{(0) \lambda}$).

With the usual definition of the propagator of an abstract discrete field $x_k$ with the action $S = x_k M^{k l} x_l / 2$ as $- i \langle x_k x_l \rangle = (M^{- 1})_{k l}$, we can write in the quasi-momentum representation in the leading order over $\Delta$
\begin{eqnarray}\label{<ww>}                                               
- i \langle w_{a b} w_{c d} \rangle & = & 2 \frac{\eta_{a c} \eta_{b d} + \eta_{a d} \eta_{b c} - \eta_{a b} \eta_{c d}}{ - \eta^{e f} \otDelta_e \tDelta_f} \nonumber \\ & - & 2 \left( 1 - \alpha^{- 1} \right) \frac{\eta_{a c} \tDelta_b \tDelta_d + \eta_{b c} \tDelta_a \tDelta_d + \eta_{a d} \tDelta_b \tDelta_c + \eta_{b d} \tDelta_a \tDelta_c}{\left( \eta^{e f} \otDelta_e \tDelta_f \right)^2} .
\end{eqnarray}

\noindent In particular, the $\Delta_\lambda \Delta_\mu$ appearing here can be treated in the leading order as a Hermitian symmetric operator, since $\Delta_\lambda \Delta_\mu - \oDelta_\mu \oDelta_\lambda = -\Delta_\lambda \oDelta_\lambda \Delta_\mu - \Delta_\lambda \Delta_\mu \oDelta_\mu - \Delta_\lambda \oDelta_\lambda \Delta_\mu \oDelta_\mu = O ( \Delta^3 )$. Or here a symmetrization of the type $\Delta_\lambda \Delta_\mu \Rightarrow - \Delta_{ ( \lambda } \oDelta_{\mu )} \equiv (\Delta_\lambda \oDelta_\mu + \Delta_\mu \oDelta_\lambda) / 2$ can be implied.

Expression (\ref{<ww>}) could follow exactly as Hermitian symmetric, if by $\Delta_\lambda$ we mean its symmetrized form $\Delta^{\rm sym}_\lambda = ( T_\lambda - \oT_\lambda ) / 2 = ( \Delta_\lambda - \oDelta_\lambda ) / 2$, which coincides with $\Delta$ in the leading order: $\Delta^{\rm sym}_\lambda = \Delta_\lambda - \Delta_\lambda \oDelta_\lambda / 2 = \Delta_\lambda + O ( \Lambda^2 )$. This form has the property $\oDelta^{\rm sym}_\lambda = - \Delta^{\rm sym}_\lambda$, the same as $\overline{\partial}_\lambda = - \partial_\lambda$, which makes it possible to transfer $\partial_\lambda$ from one factor to another one when integrating the quadratic form by parts in the continuum action. However, the disadvantage of $\Delta^{\rm sym}_\lambda$ as a finite-difference analogue of the derivative is that the result of its action on a function is weakly sensitive to changes in the function that have a phase shift between neighboring vertices close to $\pi$.

The ghost fields are rescaled to $\rtheta^a$, $\ortheta_a$ according to
\begin{equation}                                                           
\tl^{(0) \lambda}_a \rtheta^a = \vartheta^\lambda , ~~~ \tl^{(0) \lambda}_a \otheta_\lambda = \ortheta_a ,
\end{equation}

\noindent and the ghost term reads
\begin{eqnarray}                                                           
\otheta_\lambda \tilde{\cal O}^\lambda{}_\mu \vartheta^\mu & = & \ortheta_a \left\{ \tDelta_b \left[ h^{b c} \tDelta_c \rtheta^a - \left( \tDelta_c h^{a c} \right) \rtheta^b \right] + \frac{ 1 }{ 4 } \left( \tDelta_c h^{a c} \right) \tDelta_b \rtheta^b \right. \nonumber \\ & & \left. + \frac{ 1 }{ 8 } \left( \tDelta_c h^{a c} \right) \left(\tDelta_b \ln h \right) \rtheta^b \right\} .
\end{eqnarray}

\noindent The propagator of the ghost fields is
\begin{equation}\label{<theta-theta>}                                      
-i \langle \rtheta^a \ortheta_b \rangle = \frac{\delta^a_b}{ - \eta^{c d} \otDelta_c \tDelta_d} .
\end{equation}

\subsection{Features of the general expansion parameterizing the metric}\label{general}

In general notation, we can write the expansion of $\rg_{a b}$ over $\Delta \tg_{\lambda \mu}$, where all coefficients are of order at most $O(\eta^{- 1 / 2})$ (and, in addition, $(C^{- 1}_{(1)})^{ \srg_{a b}}_{\Delta \tg_{\lambda \mu}}$ is at most $O(\eta^{ 1 / 2})$), as follows from our consideration of the absence of a large parameter in the perturbative expansion in Section \ref{large-parameter},
\begin{eqnarray}\label{g=Cdg}                                              
& & \rg_{a b} = \eta_{a b} + C_{(1) \srg_{a b}}^{\Delta \tg_{\lambda \mu}} \Delta \tg_{\lambda \mu} + C_{(2) \srg_{a b}}^{\Delta \tg_{\lambda_1 \mu_1} \Delta \tg_{\lambda_2 \mu_2}} \Delta \tg_{\lambda_1 \mu_1} \Delta \tg_{\lambda_2 \mu_2} + \dots \nonumber \\ & & + C_{(n) \srg_{a b}}^{\Delta \tg_{\lambda_1 \mu_1} \dots \Delta \tg_{\lambda_n \mu_n}} \Delta \tg_{\lambda_1 \mu_1} \dots \Delta \tg_{\lambda_n \mu_n} + \dots , \nonumber \\ & & \mbox{where ~~} C_{(n) \srg_{a b}}^{\Delta \tg_{\lambda_1 \mu_1} \dots \Delta \tg_{\lambda_n \mu_n}} = O(\eta^{- 1 / 2}) ~ \forall ~ n .
\end{eqnarray}

\noindent In the variables $w_{a b} = C_{(1) \srg_{a b}}^{\Delta \tg_{\lambda \mu}} \Delta \tg_{\lambda \mu}$, this reads
\begin{eqnarray}\label{g=Bw}                                               
& & \rg_{a b} = \eta_{a b} + w_{a b} + B_{(2) \srg_{a b}}^{w_{c_1 d_1} w_{c_2 d_2}} w_{c_1 d_1} w_{c_2 d_2} + \dots \nonumber \\ & & + B_{(n) \srg_{a b}}^{w_{c_1 d_1} \dots w_{c_n d_n}} w_{c_1 d_1} \dots w_{c_n d_n} + \dots , \nonumber \\ & & \mbox{where ~~} B_{(n) \srg_{a b}}^{w_{c_1 d_1} \dots w_{c_n d_n}} = \nonumber \\ & & = C_{(n) \srg_{a b}}^{\Delta \tg_{\lambda_1 \mu_1} \dots \Delta \tg_{\lambda_n \mu_n}} (C^{- 1}_{(1)})^{ w_{c_1 d_1}}_{\Delta \tg_{\lambda_1 \mu_1}} \dots (C^{- 1}_{(1)})^{ w_{c_n d_n}}_{\Delta \tg_{\lambda_n \mu_n}} = O(\eta^{(n - 1) / 2}) .
\end{eqnarray}

\begin{figure}[h]
	\centering
	\includegraphics[scale=1]{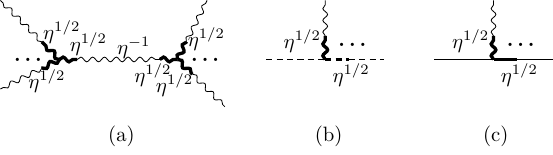}
	\caption{Maximal possible factors $\eta^{1 / 2}$ per graviton end (bold line) in the coefficients of multigraviton vertices and the value $\eta^{- 1}$ per graviton (wavy) line cancelling these factors in diagrams (a). By one factor $\eta^{1 / 2}$ smaller coefficients of multigraviton-two-ghost (b) and multigraviton-two-scalar (c) vertices with the same $\eta^{- 1}$ per ghost (dashed) or scalar (solid) line.}
	\label{f2}
\end{figure}

\noindent (The above $B$ in (\ref{g_(lm)=lg_(ab)l}) is just $B_{(2)}$.) The growing power of $\eta$ in this expansion is noteworthy. The differentiation symbol $\tDelta_a$ also contains $\sqrt{\eta}$, so in the differentiated expansion, namely the expansion of $\tDelta_c \rg_{a b}$, each $w_{c d}$ appears with the factor $\sqrt{\eta}$. Therefore, when we substitute the parametrization of $\rg_{a b}$ in terms of $w_{c d}$ into the action, the maximum degree of $\eta$ falls on those multigraviton vertices, where each graviton has an associated factor $\sqrt{\eta}$. Thus, each graviton line can contribute two factors $\sqrt{\eta}$ from its two ends. On the other hand, the propagator (\ref{<ww>}) contributes $\eta^{- 1}$ due to the overall scale of $\tDelta_a$ raised to the power of -2 (as does the ghost line (\ref{<theta-theta>}) or, without going into detail here, the scalar line below). In overall, these factors are cancelled giving $O(1)$ for the dependence on $\eta$ (Fig.~\ref{f2}(a)).

If a vertex with $n$ gravitons and two ghost fields is considered, we have maximally $n - 1$ factors $\sqrt{ \eta }$ associated with the gravitons $w_{c d}$ and two such factors contained in two $\tDelta$'s, in overall, $n + 1$ factors $\sqrt{ \eta }$ per $n + 2$ fields at the given vertex, Fig.~\ref{f2}(b). We can assume that the propagators of $n + 2$ lines converging at this vertex assign to this vertex a factor $\eta^{- 1 / 2}$ from each of them. In overall, such a vertex gives at most $O ( \eta^{- 1 / 2} )$, that is, even relatively suppressed.

We can also take a scalar field $\phi$ with mass $m$, whose continuum action is
\begin{equation}                                                           
S_\phi = - \frac{1}{2} \int \left[ g^{\lambda \mu} \left(\partial_\lambda \phi \right) \left(\partial_\mu \phi \right) + m^2 \phi^2 \right] \sqrt{- g} \d^4 x ,
\end{equation}

\noindent as the simplest example of a matter field. Its discretization is a separate task, but it is clear that in the aspect of the current discussion, the properties of the multigraviton vertices of this field will be similar to those of the ghost field, Fig.~\ref{f2}(c).

Of course, for the most part, this is simply the absence of a large parameter in the perturbative expansion considered above, expressed in other words.

To get to the continuum notation (in the leading order over metric/field variations from 4-simplex to 4-simplex), we have to go from finite differences to derivatives and from summation over sites to integration.

A shift to the neighboring site in the direction $\lambda$ in the metric $\rg^{(0)}_{a b} = \eta_{a b}$ corresponds to $\Delta x^a = \sqrt{(\eta - 10) / (3 \pi)} ( \gamma, 1, 1, 1 )$ at $a = \lambda$. That is, in the leading order, $\Delta_\lambda = l^{(0) a}_\lambda \partial_a$. Hence $\tDelta_a = \sqrt{ \det \| l^{(0) b}_\mu \|} \partial_a$.

The transition from summation to integration, $\sum_{\rm sites} ( \cdot ) \to \int ( \cdot ) \d^4 x / \Delta^4 x$, involves dividing by the coordinate volume $\Delta^4 x$ per site, that is, by $\det \| l^{(0) b}_\mu \|$. And dividing the action by $\Delta^4 x$ is equivalent to dividing each of the two $\tDelta$'s in the expression for the action (\ref{Sg[LgL^(-1)]}) by $\sqrt{ \det \| l^{(0) b}_\mu \|}$. As a result, $\tDelta_a$ is replaced by $\partial_a$, and we get the continuum action, as well as the continuum expressions for propagators known in the literature (taking into account the fact that we use the spacelike $\eta_{a b} = {\rm diag}(-1,1,1,1)$).

Limiting ourselves to the linear part of the dependence of $\rg_{a b}$ on $w_{a b}$, we formally obtain in the continuum limit the continuum perturbation theory. Conversely, if the continuum diagrams for some process and/or some structure converge (one-loop corrections to the Newtonian potential would be an interesting example), this means that these diagrams are not due to infinite momenta. Then their discrete analogs are mainly due to quasi-momenta that are not close to the limit, and for ordinary external momenta they are equal to the continuum ones with a high accuracy. But discrete diagrams are free from UV divergencies, even if their continuum counterparts are UV divergent. In this case, these discrete diagrams can serve at least for a rough estimate.

\subsection{Typical diagrams}\label{typical}

The foregoing can be illustrated by a more specific type of diagrams, for example, diagrams describing loop corrections to the graviton propagator Fig.~\ref{f2,5}. Also consider possible initial stages of evaluating such diagrams. The simplest are the one-loop diagrams Fig.~\ref{f2,5}(b,c),
\begin{figure}[h]
	\centering
	\includegraphics[scale=1]{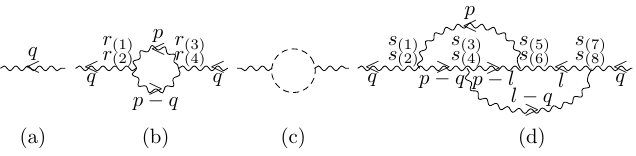}
	\caption{Some tree correlator (propagator) (a) and some one-loop (b), (c) and two-loop (d) corrections to it; $r$ or $s$ mean quasi-momenta defining finite differences entering the considered term in the corresponding interaction vertex.}
	\label{f2,5}
\end{figure}
the general term in the expressions for which has the form
\begin{eqnarray}                                                           
& & \mbox{(b,c)} \sim \int^\pi_{- \pi} \frac{\d^4 p}{( 2 \pi )^4} \frac{ \tDelta ( r_{(1) })_{a_1} \tDelta ( r_{(2) })_{a_2} \tDelta ( r_{(3) })_{a_3} \tDelta ( r_{(4) })_{a_4} }{ \frac{ \eta - 10}{3 \pi} \gamma \left( \frac{1}{\gamma^2} 4 \sin^2 \frac{p_0 }{2 } - \sum^3_{\alpha = 1} 4 \sin^2 \frac{p_\alpha }{2 } \right) } \nonumber \\ & & \cdot \frac{1}{ \frac{ \eta - 10}{3 \pi} \gamma \left( \frac{1}{\gamma^2} 4 \sin^2 \frac{ p_0 - q_0 }{2 } - \sum^3_{\alpha = 1} 4 \sin^2 \frac{p_\alpha - q_\alpha }{2 } \right) }
\end{eqnarray}

\begin{spacing}{1.37}

\noindent (thus far without the two terminal propagators), where the $\alpha$ parameter is taken equal to 1 for simplicity. Here $\pm r_{( 1 )}$, $\pm r_{( 2 )}$ and $\pm r_{( 3 )}$, $\pm r_{( 4 )}$ are pairs of any two different quasi-momenta from $p$, $p - q$, $q$ defining the finite differences $\tDelta ( r_{(j) })$ entering three-graviton vertices,
\begin{eqnarray}                                                           
& & \tDelta ( r )_a = \sqrt{\gamma \frac{\eta - 10}{3 \pi}} ( \gamma^{- 1} \Delta ( r_0 ), \Delta ( r_1 ), \Delta ( r_2 ), \Delta ( r_3 ) ), ~~~ \Delta ( r ) = \exp ( i r ) -1 , \nonumber \\ & & \overline{\Delta ( r )} = \Delta ( - r ) ,
\end{eqnarray}

\noindent and we can analyze the general term $\mbox{(b,c)} \sim $
\begin{eqnarray}\label{int-1-loop}                                      
& & \int^\pi_{- \pi} \frac{\d^4 p}{( 2 \pi )^4} \frac{ \prod^4_{j=1} \Delta ( r_{( j ) \lambda_j} )}{ \left( \sin^2 \frac{p_0 }{2 } - \sin^2 \frac{\mu_\sbp }{2 } + i0 \right) \left( \sin^2 \frac{ p_0 - q_0 }{2 } - \sin^2 \frac{\mu_{ \sbp - \sbq } }{2 } + i0 \right) } , \\ & & \label{mu-p} \sin^2 \frac{\mu_\sbp }{2 } \equiv \gamma^2 \sum^3_{\alpha = 1} \sin^2 \frac{p_\alpha }{2 } , ~~~ 0 \leq \mu_\sbp < \pi ,
\end{eqnarray}

\noindent at $\lambda_j = a_j$. Here, in order to set the direction of bypassing the propagator poles, we add an arbitrarily small imaginary value $+i0$ to its denominator (add $-i0$ to the mass squared), as in the continuum theory. Obviously, in order for the pole $\mu_\sbp$ for $p_0$ defined by formula (\ref{mu-p}) to be really real $\forall \bp = (p_1, p_2, p_3)$, the inequality
\begin{equation}\label{gamma<...}                                          
3 \gamma^2 < 1
\end{equation}

\noindent must hold. $\gamma$ was estimated based on black hole entropy calculations using the area operator spectrum in LQG in a number of papers \cite{AshBaeCorKra,LewDom,Meis,Khr} from $\gamma = ( \ln 2 ) / ( \pi \sqrt{3} ) = 0.127...$ \cite{AshBaeCorKra} to $\gamma = 0.274...$ \cite{Khr}. These values satisfy (\ref{gamma<...}) with a margin. Thus, the propagator poles are at real $p_0$ $\forall \bp$.

Consider integration with respect to $p_0$ along the contour $C$, consisting of segments $[ - \pi , + \pi ]$, $[ + \pi , + \pi +i L ]$, $[ + \pi + i L , - \pi + i L ]$, $[ - \pi +i L , - \pi ]$, $L \to \infty$, Fig.~\ref{f2,75}.
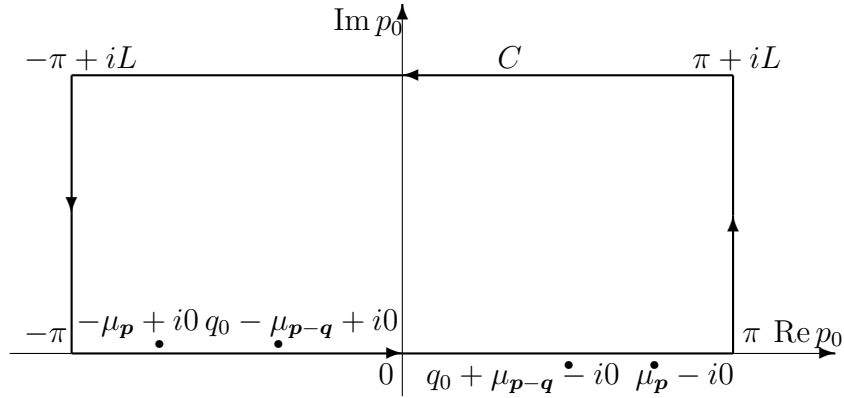
\begin{figure}[h]
\unitlength 0.9pt
\begin{picture}(342,165)(-263,-71)
\put(96,-43){\line(1,0){38}}
\put(-43,-61){\line(0,1){160}}
\put(-126,-33){$q_0 - \mu_{\sbp - \sbq} + i0$}
\put(-95,-39){\circle*{3}}
\put(-33,-56){$q_0 + \mu_{\sbp - \sbq} - i0$}
\put(27,-48){\circle*{3}}
\put(-182,-43){\line(-1,0){26}}
\put(-180,-33){$- \mu_\sbp + i0$}
\put(-202,77){$- \pi + i L$}
\put(-202,-38){$- \pi$}
\put(-72,93){$\Im p_0$}
\put(-145,-39){\circle*{3}}
\put(55,-56){$\mu_\sbp - i0$}
\put(63,-48){\circle*{3}}
\put(-53,-56){$0$}
\put(79,77){$\pi + i L$}
\put(-3,77){$C$}
\put(100,-38){$\pi$}
\put(113,-38){$\Re p_0$}

\thicklines

\put(134,-43){\vector(1,0){5}}
\put(-43,99){\vector(0,1){5}}
\put(96,-43){\vector(0,1){58}}
\put(96,15){\line(0,1){59}}
\put(96,74){\vector(-1,0){139}}
\put(-43,74){\line(-1,0){139}}
\put(-182,74){\vector(0,-1){58}}
\put(-182,16){\line(0,-1){59}}
\put(-182,-43){\vector(1,0){139}}
\put(-43,-43){\line(1,0){139}}

\end{picture}
\caption{Integration contour in the plane of complex $p_0$.
\label{f2,75}}
\end{figure}
On the segments $[ + \pi , + \pi +i L ]$ and $[- \pi, - \pi +i L ]$, the values of the integrand coincide because of its periodicity in $\Re p_0$, while these segments are traversed in opposite directions. Therefore, their contributions are mutually cancelled. Thus, the integral of interest is the sum of the contour integral and the integral over the remote segment $[ - \pi + i L , + \pi + i L ]$, $\int^\pi_{- \pi} = \oint_C + \int^{ \pi + i L}_{- \pi + i L}$; in turn, $\oint_C$ is equal to the sum of the residues at the poles inside the contour.

\end{spacing}

Consider first the integral over the remote segment $[ - \pi + i L , + \pi + i L ]$. If $p_0 = x + iL$, then the denominator in (\ref{int-1-loop}) behaves as $\exp (2L - 2ix + iq_0)$. In the numerator $\prod^4_{j=1} \Delta ( r_{( j ) \lambda_j} )$, along with $\Delta ( p )$, $\Delta ( p - q )$ tending to zero at $L \to \infty$, due to the Hermitian symmetrization there are also $\Delta ( - p ) \Rightarrow \exp ( L - ix )$ and $\Delta ( - p + q ) \Rightarrow \exp ( L - ix + iq )$. In principle, terms that are maximally proportional to $\exp ( 4 L )$ are possible, which, when divided by the denominator, give proportionality to $\exp ( 2 L )$. However, each $\exp L$ is accompanied by a factor $\exp ( - i x )$, so that the terms of the integrand growing at $L \to \infty$ vanish when integrating over $x \in [- \pi, \pi ]$. A nonzero result when integrating over $x$ is given only by terms that are constant at $L \to \infty$ and, therefore, finite, $\propto \exp ( 0\cdot L )$. So a priori we will end up with a linear combination of a constant with $\exp ( i q )$ and $\exp ( - i q )$ (as in the subsequent finite integration over $\d^3 \bp$), with constants $A_0$, $A_+$, $A_-$ depending on the actual choice of $r_{( j )}$ and their indices $\lambda_j$,
\begin{equation}                                                           
\int^{\pi + iL}_{- \pi + iL} \frac{\d p_0}{ 2 \pi } \frac{ \prod^4_{j=1} \Delta ( r_{( j ) \lambda_j} )}{ \left( \sin^2 \frac{p_0 }{2 } - \sin^2 \frac{\mu_\sbp }{2 } \right) \left( \sin^2 \frac{ p_0 - q_0 }{2 } - \sin^2 \frac{\mu_{ \sbp - \sbq } }{2 } \right) } = A_0 + A_+ e^{i q} + A_- e^{-i q} .
\end{equation}
A continuum analogue would be integration over $\d \tp_0$ over a large semicircle in the upper (lower) half-plane of the complex $\tp_0$ in
\begin{equation}\label{1-loop-cont}                                        
\int \frac{\d^4 \tp}{( 2 \pi )^4} \frac{ \tr_{(1) \lambda_1} \tr_{(2) \lambda_2} \tr_{(3) \lambda_3} \tr_{(4) \lambda_4} }{(\tp_0^2 - \tbp^2) [ (\tp_0 - \tq_0 )^2 - (\tbp - \tbq )^2]}, ~~~ \tr_{(n)} = \tp \mbox{ or } \tp - \tq
\end{equation}

\noindent (here the tilde means momentum, to distinguish it from quasi-momentum), which is infinite and, together with subsequent integration over $\d^3 \tbp$, gives a fourth-order polynomial in $\tq$ with infinite (depending on UV cutoff) coefficients.

More interesting is the sum of the residues at the poles $- \mu_\sbp + i0$ and $q_0 - \mu_{\sbp - \sbq} + i0$, which we estimate for a sort of master integral as
\begin{eqnarray}\label{int-master}                                         
& & \frac{1}{i} \oint_C \frac{\d p_0}{ 2 \pi } \frac{ 1 }{ \left( \sin^2 \frac{p_0 }{2 } - \sin^2 \frac{\mu_\sbp }{2 } + i0 \right) \left( \sin^2 \frac{ p_0 - q_0 }{2 } - \sin^2 \frac{\mu_{ \sbp - \sbq } }{2 } + i0 \right) } \nonumber \\ & & = \frac{1}{\sin \mu_\sbp } \cdot \frac{2}{\sin^2 \frac{\mu_\sbp - q_0 }{2 } - \sin^2 \frac{\mu_{\sbp - \sbq}}{2} } + \frac{1}{\sin \mu_{\sbp - \sbq} } \cdot \frac{2}{\sin^2 \frac{\mu_{\sbp - \sbq } + q_0 }{2 } - \sin^2 \frac{\mu_\sbp}{2} } \nonumber \\ & & \stackrel{\mbox{\scriptsize $\mu \to \mu - i0$}}{ \longrightarrow } \frac{\sin (\mu_\sbp + \mu_{\sbp - \sbq})}{\sin \mu_\sbp \sin \mu_{\sbp - \sbq}} \cdot \frac{2 }{\sin^2 \frac{q_0}{2 } - \sin^2 \frac{\mu_\sbp + \mu_{\sbp - \sbq}}{2 } + i0 \cdot {\rm sgn} \sin (\mu_\sbp + \mu_{\sbp - \sbq})} .
\end{eqnarray}

\noindent The real part of the subsequent integral over $\d^3 \bp$ is defined in the sense of the Cauchy principal value, plus we can take the delta-function of the denominator to which $i0$ is assigned to find its imaginary part (the Sokhotski–Plemelj formula), similarly as in the continuum theory, but there are no UV divergences. Analogously, we can integrate over the 0-component of the loop quasi-momentum in each loop, say, over $\d p_0 \d l_0$ for the diagram Fig.~\ref{f2,5}(d).

If the contribution to the diagram is provided by sufficiently small quasi-momenta, then their sines can be replaced by their arguments themselves. Then we can go from any quasi-momentum $p$ to the corresponding momentum $\tp$ using the above timelike $ b_{\rm t} $ and spacelike $ b_{\rm s} $ length scales:
\begin{equation}                                                           
p_0 = b_{\rm t} \tp_0 = \gamma b_{\rm s} \tp_0 , ~~~ p_\alpha = b_{\rm s} \tp_\alpha , ~~~ \d^4 p = \gamma \left( \frac{ \eta - 10 }{3 \pi} \right)^2 \d^4 \tp = \sqrt{- g^{(0 )}} \d^4 \tp .
\end{equation}

\noindent Fig.~\ref{f2,5}(a) leads to $\sim i / ( \tq^2 \sqrt{- g^{(0 )}} )$. For Fig.~\ref{f2,5}(b,c), the result of this transition is proportional to the integral (\ref{1-loop-cont}) times $\sqrt{- g^{(0 )}}$. Each of the two propagators and each of the two interaction vertices contributes an imaginary unit $i$ to the factor, $i^4=1$ in total. In addition, loop integration (\ref{int-master}) gives $i$, and in the full expression, including two terminal propagators, each of these propagators contributes $\sim i / ( \tq^2 \sqrt{- g^{(0 )}} )$ to the factor. For the diagram in Fig.~\ref{f2,5}(d), such a full expression, up to the assumption of smallness of the quasi-momenta, has the form
\begin{eqnarray}                                                           
& & I_{\rm ( d )} \sim \frac{i^2}{\left( \frac{\eta - 10}{3 \pi} \gamma \right)^2 \left( \frac{1}{\gamma^2} \sin^2 \frac{q_0 }{2 } - \sum^3_{\alpha = 1} \sin^2 \frac{q_\alpha }{2 } \right)^2} \int\limits^\pi_{- \pi} \frac{\d^4 p}{(2 \pi)^4} \frac{\d^4 l}{(2 \pi)^4} \nonumber \\ & & \cdot \frac{i^{5+ 4} \prod^8_{j = 1} \tDelta ( s_{( j )})_{b_j}}{\left( \frac{\eta - 10}{3 \pi} \gamma \right)^5 \left( \frac{1}{\gamma^2} \sin^2 \frac{ l_0 - q_0 }{2 } - \sum^3_{\alpha = 1} \sin^2 \frac{l_\alpha - q_\alpha }{2 } \right) \left( \frac{1}{\gamma^2} \sin^2 \frac{ p_0 - q_0 }{2 } - \sum^3_{\alpha = 1} \sin^2 \frac{p_\alpha - q_\alpha }{2 } \right)} \nonumber \\ & & \cdot \frac{1 }{ \left( \frac{1}{\gamma^2} \sin^2 \frac{l_0 }{2 } - \sum^3_{\alpha = 1} \sin^2 \frac{l_\alpha }{2 } \right) \left( \frac{1}{\gamma^2} \sin^2 \frac{p_0 }{2 } - \sum^3_{\alpha = 1} \sin^2 \frac{p_\alpha }{2 } \right) } \nonumber \\ & & \cdot \frac{1 }{ \left( \frac{1}{\gamma^2} \sin^2 \frac{ p_0 - l_0 }{2 } - \sum^3_{\alpha = 1} \sin^2 \frac{p_\alpha - l_\alpha }{2 } \right) } .
\end{eqnarray}

\noindent Here $( \pm s_{(1 )}, \pm s_{(2 )})$, $( \pm s_{(3 )}, \pm s_{(4 )})$, $( \pm s_{(5 )}, \pm s_{(6 )})$, and $( \pm s_{(7 )}, \pm s_{(8 )})$ are pairs of any two different quasi-momenta from $(q, p, p - q )$, $(p - l, p - q, l - q )$, $(l, p, p - l )$, and $(q, l, l - q )$, respectively. After transition from quasi-momenta to momenta, this in total against the background of the terms Fig.~\ref{f2,5}(a,b,c) looks like
\begin{eqnarray}                                                          
& & I_{\rm (a)} + I_{\rm (b,c)} + I_{\rm (d)} = \frac{{\rm const} \cdot i}{\sqrt{- g^{ (0)}} \tq^2} + \frac{{\rm const} }{\sqrt{- g^{ (0)}} ( \tq^2 )^2 } \int \frac{\d^4 \tp }{ (2 \pi )^4 } \frac{\prod^4_{j = 1} \tr_{(j ) \lambda_j } }{\tp^2 (\tp - \tq )^2 } \nonumber \\ & & + \frac{{\rm const} \cdot i}{\sqrt{- g^{ (0)}} ( \tq^2 )^2 } \int \frac{\d^4 \tp }{ (2 \pi )^4 } \frac{\d^4 \tl }{ (2 \pi )^4 } \frac{\prod^8_{j = 1} \ts_{(j) \mu_j}}{(\tl - \tq)^2 (\tp - \tq)^2 \tl^2 \tp^2 (\tp - \tl )^2} .
\end{eqnarray}

\noindent Here the (real) constants depend on the external indices and on the actual choice of $r_{( j )}$, $s_{( j )}$ and their indices $\lambda_j$, $\mu_j$.

\begin{figure}[h]
	\centering
	\includegraphics[scale=1]{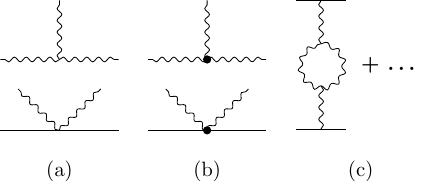}
	\caption{The (discrete form of the) standard vertices (a) and our additional ones (b) that contribute to the Newtonian law on the one-loop level (c).}
	\label{f3}
\end{figure}

New contributions appear when going beyond the linear part of the dependence of $\rg_{a b}$ on $w_{a b}$. In particular, again in connection with one-loop corrections to the Newtonian potential, in addition to the usual three-graviton and two-graviton-two-matter vertices (Fig.~\ref{f3}(a)), we can encounter a three-graviton and two-graviton-two-matter vertices (Fig.~\ref{f3}(b)) arising from the $w_{c d}$-bilinear term in $\rg_{a b}$. On the one-loop level for the corrections to the Newtonian potential (Fig.~\ref{f3}(c)), these vertices could be taken into account; since the $w_{c d}$-bilinear term in $\rg_{a b}$ has an additional factor $\sqrt{\eta}$, the contribution of the additional vertices can be significant.

\subsection{Dual construction of the metric parametrization}\label{dual}

The parametrization $g_{\lambda \mu} = g_{\lambda \mu} ( \{ \tg_{\nu \rho} \} )$ also can be rewritten in the variables $h^{\lambda \mu}$: $h^{\lambda \mu} = h^{\lambda \mu} ( \{ \tih^{\nu \rho} \} )$, where $\tih^{\lambda \mu}$ is defined in terms of $\tg_{\nu \rho}$ just as $h^{\lambda \mu}$ is defined in terms of $g_{\nu \rho}$. This is some change of variables, which, as it were, does not greatly simplify the system under consideration, but rather the opposite. This is because the procedure of transforming the measure was determined by the dependence of its arguments $\{ \rmm_\alpha , \rt_\alpha \}$ on $g_{\lambda \mu}$ (\ref{mt=f(g)}). So far, we have used the spacelike areas $\{ \rmm_\alpha \}$ to express in terms of them the covariant components of the metric $\{ g_{\alpha \alpha} \}$; after that, $ g_{0 0} $ is determined by one variable of the type of a timelike area, whose role is played by the sum $\sum_\alpha \rt_\alpha$. The components $g_{\lambda \no \mu}$ define corrections. This procedure is natural for the variables $g_{\lambda \mu}$ and unnatural for the variables $h^{\lambda \mu}$, in which the areas take the form
\begin{eqnarray}                                                           
& & \rmm_\alpha = \sqrt{ - h^{0 0} h^{\alpha \alpha} + (h^{0 \alpha})^2} , ~~~ \rt_1 = \sqrt{ h^{2 2} h^{3 3} - ( h^{2 3})^2}, ~ 2 ~ \mbox{perm}(123) , \nonumber \\ & & \rmm \stackrel{\rm def}{=} \sum_\alpha \rmm_\alpha .
\end{eqnarray}

\noindent But here it is appropriate to apply a similar method to the permuted variables $\rmm_\alpha$ and $\rt_\alpha$ (more accurately, $\rmm_\alpha \leftrightarrow \rt_\alpha / \gamma$ in the exponent). Namely, $\{ \rt_\alpha \}$ can be used to find $\{ h^{\alpha \alpha} \}$, then the combined variable $\rmm = \sum_\alpha \rmm_\alpha$ allows to express $ h^{0 0} $. The components $h^{\lambda \no \mu}$ define corrections.

Such a formula, thus obtained for the parametrization $h^{\lambda \mu} = h^{\lambda \mu} ( \{ \tih^{\nu \rho} \} )$, must follow from $g_{\lambda \mu} = g_{\lambda \mu} ( \{ \tg_{\nu \rho} \} )$ by replacing $g_{\lambda \mu}$ with $h^{\lambda \mu}$ and  $\tg_{\lambda \mu}$ with $\tih^{\lambda \mu}$ and by slightly adjusting the dependence on $\gamma$, which is simple (general power coefficient in the expression for the tensor component).

When analogously to the case of $g_{\lambda \mu}$ rescaling the coordinate axes, the newly introduced in (\ref{Sg[LgL^(-1)]}) densitized tetrad $\tl^{(0) \lambda}_a$ appears,
\begin{eqnarray}\label{h^(lm)=lh^(ab)l}                                    
& & h^{\lambda \mu} = \tl_a^{(0) \lambda} \rh^{a b} \tl_b^{(0) \mu}, ~~~ \tl^{(0) \lambda}_a = \sqrt{\frac{\eta - 10}{3 \pi}} {\rm diag}(\gamma^{-1 / 2},\gamma^{1 / 2},\gamma^{1 / 2},\gamma^{1 / 2}),
\end{eqnarray}

\noindent where, in terms of $\Delta \tih^{\lambda \mu} = \tih^{\lambda \mu} - \tih^{ ( 0 ) \lambda \mu} = \tih^{\lambda \mu} - \eta^{\lambda \mu} $,
\begin{eqnarray}                                                           
& & \rh^{a b} = \eta^{a b} + C_{(1) \srh^{a b}}^{\Delta \tih^{\lambda \mu}} \Delta \tih^{\lambda \mu} + C_{(2) \srh^{a b}}^{\Delta \tih^{\lambda_1 \mu_1} \Delta \tih^{\lambda_2 \mu_2}} \Delta \tih^{\lambda_1 \mu_1} \Delta \tih^{\lambda_2 \mu_2} + \dots \nonumber \\ & & + C_{(n) \srh^{a b}}^{\Delta \tih^{\lambda_1 \mu_1} \dots \Delta \tih^{\lambda_n \mu_n}} \Delta \tih^{\lambda_1 \mu_1} \dots \Delta \tih^{\lambda_n \mu_n} + \dots .
\end{eqnarray}

\noindent Linear transformations of $\tih^{\lambda \mu}$ as parametrization variables are allowed, since they do not change, up to proportionality, the form of the measure $\d^{10} \tih^{\lambda \mu}$. In the variables $u^{a b} = C_{(1) \srh^{a b}}^{\Delta \tih^{\lambda \mu}} \Delta \tih^{\lambda \mu}$, this reads
\begin{eqnarray}                                                           
& & \rh^{a b} = \eta^{a b} + u^{a b} + B_{(2) \srh^{a b}}^{u^{c_1 d_1} u^{c_2 d_2}} u^{c_1 d_1} u^{c_2 d_2} + \dots \nonumber \\ & & + B_{(n) \srh^{a b}}^{u^{c_1 d_1} \dots u^{c_n d_n}} u^{c_1 d_1} \dots u^{c_n d_n} + \dots , \nonumber \\ & & \mbox{where ~~} B_{(n) \srh^{a b}}^{u^{c_1 d_1} \dots u^{c_n d_n}} = C_{(n) \srh^{a b}}^{\Delta \tih^{\lambda_1 \mu_1} \dots \Delta \tih^{\lambda_n \mu_n}} (C^{- 1}_{(1)})^{ u^{c_1 d_1}}_{\Delta \tih^{\lambda_1 \mu_1}} \dots (C^{- 1}_{(1)})^{ u^{c_n d_n}}_{\Delta \tih^{\lambda_n \mu_n}}.
\end{eqnarray}

\noindent We have $C_{(n) \srh^{a b}}^{\Delta \tih^{\lambda_1 \mu_1} \dots \Delta \tih^{\lambda_n \mu_n}} = C_{(n) \srg_{a b}}^{\Delta \tg_{\lambda_1 \mu_1} \dots \Delta \tg_{\lambda_n \mu_n}}$, $B_{(n) \srh^{a b}}^{u^{c_1 d_1} \dots u^{c_n d_n}} = B_{(n) \srg_{a b}}^{w_{c_1 d_1} \dots w_{c_n d_n}}$. Correspondingly, the properties of the considered expansion are the same as in the case of $g_{\lambda \mu}$, eqs. (\ref{g=Cdg}), (\ref{g=Bw}). $h^{\lambda \mu} = h^{\lambda \mu} ( \{ \tih^{\nu \rho} \} )$, up to bilinear terms, is given in \ref{appendix1} (cf. (\ref{g(lm)=1+Dtilde(g)})). For the dependence of $h^{\lambda \mu}$ on $u^{a b}$ up to bilinear terms and in the leading order over $\eta$ we have (cf. (\ref{g_(ab)}))

\begin{eqnarray}                                                           
& & \hspace{-5mm} \frac{3 \pi \gamma}{ \eta - 10} h^{0 0} = - 1 + u^{0 0} + \frac{\sqrt{6}}{5} \sqrt{\frac{\eta - 12}{\pi }} \left\{ - \frac{\sqrt{3}}{4} \sqrt{\frac{\eta - 10}{\eta - 12}} \left( u^{0 0} \right)^2 + \frac{\sqrt{3} + 2}{6} u^{0 0} \sum_\alpha u^{\alpha \alpha} \right. \nonumber \\ & & \left. + \sqrt{\frac{\eta - 12}{\eta - 10}} \left[ \frac{4 \sqrt{3} - 3}{72} \left( \sum_\alpha u^{\alpha \alpha} \right)^2 + \frac{6 \sqrt{3} - 7}{72} \sum_\alpha \left( u^{\alpha \alpha} \right)^2 \right] \right\} , \nonumber \\ & & \frac{3 \pi \gamma^{-1} }{ \eta - 10} h^{1 1} = 1 + h^{1 1} + \frac{\sqrt{6}}{5} \sqrt{\frac{\eta - 12}{\pi }} \left[ - \frac{1}{12} \left( u^{1 1} \right)^2 + \frac{5}{12} \left( u^{3 3} u^{1 1} + u^{1 1} u^{2 2} \right. \right. \nonumber \\ & & \left. \left. - u^{2 2} u^{3 3} \right) \vphantom{\frac{5}{12 }} \right], ~ 2 ~ \mbox{perm}(123), \nonumber \\ & & \frac{3 \pi }{ \eta - 10} h^{0 \alpha} = u^{0 \alpha} + \frac{\sqrt{6}}{10} \sqrt{\frac{\eta - 12}{\pi }} \left[ \sqrt{\frac{\eta - 10}{3 (\eta - 12)}} \left( u^{0 0} - \frac{1}{3} \sum_\beta u^{\beta \beta} \right) - u^{\alpha \alpha} \right. \nonumber \\ & & \left. + \frac{1}{3}  \sum_\beta u^{\beta \beta} \right] u^{0 \alpha} , \nonumber \\ & & \frac{3 \pi \gamma^{- 1} }{ \eta - 10} h^{\alpha \beta} = u^{\alpha \beta} - \frac{\sqrt{6}}{10} \sqrt{\frac{\eta - 12}{\pi }} \left( u^{\alpha \alpha} + u^{\beta \beta} \right) u^{\alpha \beta}, ~~~ \alpha \neq \beta .
\end{eqnarray}

Substituting $h^{\lambda \mu}$ in terms of $\rh^{a b}$ (\ref{h^(lm)=lh^(ab)l}) into the action (\ref{Sg[g](h)}) gives
\begin{eqnarray}                                                           
\tilS_{\rm g} [ g ] & = & \frac{1}{ 8 } \sum_{\rm sites} [ - 2 \rh^{a c} ( \tDelta_a \rh^{b d} ) ( \tDelta_d \rh_{b c} ) + \rh^{a d} ( \tDelta_a \rh^{b c} ) ( \tDelta_d h_{b c} ) \nonumber \\ & & + \frac{1}{2} \rh^{a b} ( \tDelta_a \ln \rh ) ( \tDelta_b \ln \rh )] .
\end{eqnarray}

The $\alpha$-term with the chosen $\teta_{\lambda \mu}$ (\ref{tilde-eta}) is the same as in (\ref{alpha-term-g}), with the difference that $\rh^{a b}$ is parameterized now not by $w_{c d}$ through $\rg_{e f}$, but immediately (by $u^{c d}$),
\begin{equation}                                                           
\cF [ g ] = - \frac{\alpha }{4 } \sum_{\rm sites} \left(\tDelta_c \rh^{a c} \right) (- \rh)^{- 1 / 4} \eta_{a b} \left(\tDelta_d \rh^{b d} \right) .
\end{equation}

The propagator is
\begin{eqnarray}                                                           
& & - i \langle u^{a b} u^{c d} \rangle = 2 \frac{\eta^{a c} \eta^{b d} + \eta^{a d} \eta^{b c} + (\alpha^{- 1} - 2) \eta^{a b} \eta^{c d}}{ - \eta^{e f} \otDelta_e \tDelta_f} + 2 \left( 1 - \alpha^{- 1} \right) \nonumber \\ & & \cdot \frac{2\eta^{a b} \tDelta^c \tDelta^d + 2 \eta^{c d} \tDelta^a \tDelta^b - \eta^{a c} \tDelta^b \tDelta^d - \eta^{b c} \tDelta^a \tDelta^d - \eta^{a d} \tDelta^b \tDelta^c - \eta^{b d} \tDelta^a \tDelta^c}{\left( \eta^{e f} \otDelta_e \tDelta_f \right)^2} .
\end{eqnarray}

\section{Conclusion}

The considered perturbative expansion contains the continuum one as a part, more exactly, converging continuum diagrams (for example, one-loop corrections to the Newtonian potential) are reproduced by their discrete counterparts (for habitual external momenta or distances much larger than the typical length scale (\ref{l=l-pl-sqrt-eta})). Of course, discrete diagrams are free from UV divergencies, even if their continuum counterparts are UV divergent, and they can be used for rough estimates.

Along with this, there are extra multigraviton vertices (including multigraviton-ghost and multigraviton-matter ones) and, accordingly, extra diagrams.

The discrete diagrams whose continuum counterparts converge are mostly contributed by non-maximal quasi-momenta or by virtual configurations in which we can confine ourselves to the leading order over metric variations from 4-simplex to 4-simplex. In this order,  the action is symmetric with respect to the finite-difference form of the diffeomorphism transformation (\ref{delta-g=xi}). The symmetry of the complete functional integral, including the measure, is spontaneously broken by the choice of a certain simplicial structure, in the simplest version, the orientation of the axes of the hypercubic lattice in our case. The RC strategy implies summation/averaging over all possible simplicial structures, and this should restore the symmetry. In particular, this should lead to independence from the gauge parameters, here from $\alpha$, in the leading order over metric variations from 4-simplex to 4-simplex.

The simplest such averaging can be averaging over the orientation of the hypercubic axes or (in the case of the Newtonian potential) averaging over all orientations of a pair of interacting bodies with respect to the hypercubic axes.

We have not considered a bare cosmological constant, thus avoiding expansion around a curved background.

It is important that the perturbative expansion does not contain increasing powers of $\eta$ (we recall that this is a large parameter).

And this is for the choice of the initial point of the perturbative expansion sufficiently close to the maximum point of the functional measure. Such a choice was based on minimizing the determinant (\ref{def-l0}) of the second-order form (\ref{ddS}) in the exponent in order to maximize the contribution of this form. Now this choice seems to have a more advanced dynamic justification: with this choice, the perturbative expansion does not contain increasing powers of $\eta$. These powers of $\eta$ arise if the initial point of the perturbative expansion is taken far enough from the maximum point of the measure, and the perturbative expansion explodes.

Were it not for the additional vertices, we would have a "naive" discretization of gravity, when the diagrams are the same as in the continuum GR, and only a lattice cutoff is imposed. This leads to negative powers of the length scale $l$ (\ref{l=l-pl-sqrt-eta}). They are made dimensionless by the gravity coupling, that is, by positive powers of $l_{\rm Pl}$. This results in an expansion in the small parameter $( l_{\rm Pl} / l )^2 \sim \eta^{- 1} $, and the theory would be good like QED. However, the $l$ fixing mechanism leads to extra vertices with extra positive powers of $\eta$; fortunately, at least, there is no large parameter, and the expansion proceeds in loop smallness, as at the usual ($l = l_{\rm Pl}$) cutoff.

The effects associated with the non-simple form of the functional measure flow into the parametrization of the metric, $g_{\lambda \mu} = g_{\lambda \mu} ( \{ \tg_{\nu \rho} \} )$. We have explicitly found $g_{\lambda \mu} = g_{\lambda \mu} ( \{ \tg_{\nu \rho} \} )$ up to bilinear terms over $\tg_{\lambda \mu}$. These terms lead to additional three-graviton and two-graviton-two-matter vertices and corresponding additional diagrams, in particular, they can contribute to one-loop corrections to the Newtonian potential.

We can use the contravariant density $h^{\lambda \mu} = g^{\lambda \mu} \sqrt{ - g }$ instead of $g_{\lambda \mu}$ from the very beginning (the measure is symmetric with respect to the replacement of $\{ g_{\lambda \mu} \}$ by $\{ h^{\lambda \mu} \}$, which simplifies the problem). This gives the dual parametrization $h^{\lambda \mu} = h^{\lambda \mu} ( \{ \tih^{\nu \rho} \} )$ and other diagram technique in terms of $h^{\lambda \mu}$.

The poles of propagators generally may not be at real energy (0-component of momenta) depending on the ratio $\gamma$ between the timelike $ b_{\rm t} $ and spacelike $ b_{\rm s} $ edge length scales. It is interesting that the existing estimates of $\gamma$ based on black hole entropy calculations satisfy the condition (\ref{gamma<...}) required for that the propagator poles be at real quasi-energy for any spatial quasi-momenta components.

\section*{Acknowledgments}

The present work was supported by the Ministry of Education and Science of the Russian Federation.

\appendix

\section{$h^{\lambda \mu} = h^{\lambda \mu} ( \{ \tih^{\nu \rho} \} )$ up to bilinear terms}\label{appendix1}

\begin{eqnarray}
& & \frac{3 \pi \gamma}{ \eta - 10} h^{0 0} = - 1 + 5 \sqrt{ \frac{ \pi }{2 (\eta - 10)}} \left\{ \Delta \tih^{0 0} - \frac{1}{3} \left(1 - \sqrt{\frac{\eta - 10}{3 (\eta - 12)}}  \right) \sum_\alpha \Delta \tih^{\alpha \alpha} \right. \nonumber \\ & & \left. - \frac{3}{4} \left[ \Delta \tih^{0 0} - \frac{1}{3} \left(1 - \sqrt{\frac{\eta - 10}{3 (\eta - 12)}}  \right) \sum_\alpha \Delta \tih^{\alpha \alpha} \right]^2 + \left(\frac{1 }{3 } + \frac{1 }{2 } \sqrt{\frac{\eta - 10}{3 (\eta - 12)}} \right) \right. \nonumber \\ & & \left. \cdot \left[ \Delta \tih^{0 0} - \frac{1}{3} \left(1 - \sqrt{\frac{\eta - 10}{3 (\eta - 12)}} \right) \sum_\alpha \Delta \tih^{\alpha \alpha} \right] \sum_\alpha \Delta \tih^{\alpha \alpha} + \frac{1}{6} \left( \frac{\eta - 13}{3 (\eta - 12)} \right. \right. \nonumber \\ & & \left. \left. - \frac{1}{4 } \sqrt{\frac{\eta - 10}{3 (\eta - 12)}} \right) \left(\sum_\alpha \Delta \tih^{\alpha \alpha} \right)^2 + \left(\frac{1 }{12 } - \frac{7 }{72 } \sqrt{\frac{\eta - 10}{3 (\eta - 12)}} \right) \sum_\alpha \left( \Delta \tih^{\alpha \alpha} \right)^2 \right\} \nonumber \\ & & + \frac{1 }{3 } \left( 1 - 5 \sqrt{ \frac{ \pi }{2 (\eta - 10)}} \right) \sum_\alpha \left( \tih^{\hnul \halpha} \right)^2 + \frac{1 }{6 } \left( 1 - 5 \sqrt{ \frac{ \pi }{6 (\eta - 12)}} \right) \sum_{\alpha \no \beta} \left( \tih^{\halpha \hbeta} \right)^2 \nonumber \\ & & + \frac{25 }{216} \frac{\pi }{\eta - 12} \left[ 2 \left( \sum_\alpha \Delta \tih^{\alpha \alpha} \right)^2 + \sum_\alpha \left( \Delta \tih^{\alpha \alpha} \right)^2 \right] , \nonumber \\ & & \frac{3 \pi \gamma^{- 1} }{ \eta - 10} h^{1 1} = 1 + 5 \sqrt{ \frac{ \pi }{6 (\eta - 12)}} \left[ \Delta \tih^{1 1} - \frac{1}{12} \left( \Delta \tih^{1 1} \right)^2 + \frac{5}{12} \left( \Delta \tih^{3 3} \tih^{1 1} + \Delta \tih^{1 1} \tih^{2 2} \right. \right. \nonumber \\ & & \left. \left. - \Delta \tih^{2 2} \tih^{3 3} \right) \vphantom{\frac{5 }{12 }} \right] + \left[ \frac{1}{2} - \frac{5}{2 } \sqrt{ \frac{ \pi }{6 (\eta - 12)}} \right] \left( \left( \tih^{\hthr \hone} \right)^2 + \left( \tih^{\hone \htwo} \right)^2 - \left( \tih^{\htwo \hthr} \right)^2 \right) + \frac{25 }{72 } \frac{\pi }{ \eta - 12 } \nonumber \\ & & \cdot \left[ \left( \Delta \tih^{33} + \Delta \tih^{11} \right)^2 + \left( \Delta \tih^{11} + \Delta \tih^{22} \right)^2 + 2 \left( \Delta \tih^{22} + \Delta \tih^{33} \right)^2 \right] , ~ 2 ~ \mbox{perm}(123), \nonumber \\ & & \frac{3 \pi }{ \eta - 10} h^{0 \alpha} = \tih^{0 \alpha} + \frac{1}{2} \left[ \Delta \tih^{0 0} - \Delta \tih^{\alpha \alpha} - 5 \sqrt{\frac{\pi }{2 (\eta - 10)}} \left( \Delta \tih^{0 0} - \frac{1}{3} \sum_\beta \Delta \tih^{\beta \beta} \right) \right. \nonumber \\ & & \left.+ 5 \sqrt{\frac{\pi }{6 (\eta - 12)}} \left( \Delta \tih^{\alpha \alpha} - \frac{1}{3} \sum_\beta \Delta \tih^{\beta \beta} \right) \right] \tih^{0 \alpha} , \nonumber \\ & & \frac{3 \pi \gamma^{- 1}}{ \eta - 10} h^{\alpha \beta} = \tih^{\alpha \beta} - \frac{1}{2} \left( 1 - 5 \sqrt{ \frac{ \pi }{6 (\eta - 12)}} \right) \left( \Delta \tih^{\alpha \alpha} + \Delta \tih^{\beta \beta} \right) \tih^{\alpha \beta}, ~~~ \alpha \neq \beta .
\end{eqnarray}

\end{document}